\documentclass[journal=jctcce,manuscript=article,usetitle=true]{achemso}
\usepackage[version=4]{mhchem}
\usepackage[english]{babel}
\usepackage{amsmath}
\usepackage{caption}
\usepackage{braket}
\usepackage{parskip}
\usepackage{graphicx}
\usepackage{setspace}
\usepackage{float}
\usepackage{multirow}
\usepackage{wrapfig}
\usepackage{csvsimple}
\usepackage[toc,page]{appendix}
\usepackage{caption}
\usepackage[table,xcdraw]{xcolor}
\usepackage{xcolor,colortbl}
\usepackage{titlesec}
\usepackage{hhline}
\usepackage{longtable}
\usepackage{booktabs}
\usepackage[normalem]{ulem}
\usepackage{letltxmacro}

\LetLtxMacro{\ORIGselectlanguage}{\selectlanguage}
\makeatletter
\DeclareRobustCommand{\selectlanguage}[1]{%
	\@ifundefined{alias@\string#1}
	{\ORIGselectlanguage{#1}}
	{\begingroup\edef\x{\endgroup
			\noexpand\ORIGselectlanguage{\@nameuse{alias@#1}}}\x}%
}
\newcommand{\definelanguagealias}[2]{%
	\@namedef{alias@#1}{#2}%
}
\makeatother
\definelanguagealias{en}{english}
\definelanguagealias{EN}{english}
\newcommand{\aetbigfigsize}{7in}
\newcommand{\aetfigsize}{8.45cm}
\title{A comparison of the Bravyi-Kitaev and Jordan-Wigner transformations for the quantum simulation of quantum chemistry}
\author{Andrew Tranter}
\affiliation{Department of Physics, Imperial College London, London, SW7 2AZ, United Kingdom}
\affiliation{Centre for Computational Science, University College London, London, WC1H 0AJ, United Kingdom}
\email{a.tranter13@imperial.ac.uk} 
\author{Peter J. Love}
\affiliation{Department of Physics, Tufts University, Medford, Massachusetts 02155, USA}
\author{Florian Mintert}
\affiliation{Department of Physics, Imperial College London, London, SW7 2AZ, United Kingdom}
\author{Peter V. Coveney}
\affiliation{Centre for Computational Science, University College London, London, WC1H 0AJ, United Kingdom}
\email{p.v.coveney@ucl.ac.uk}
\begin{document}
	\begin{center}
		\today
	\end{center}
	\begin{abstract}
		The ability to perform classically intractable electronic structure calculations is often cited as one of the principal applications of quantum computing.  A great deal of theoretical algorithmic development has been performed in support of this goal.  Most techniques require a scheme for mapping electronic states and operations to states of and operations upon qubits.  The two most commonly used techniques for this are the Jordan-Wigner transformation and the Bravyi-Kitaev transformation.  However, comparisons of these schemes have previously been limited to individual small molecules.  In this paper we discuss resource implications for the use of the Bravyi-Kitaev mapping scheme, specifically with regard to the number of quantum gates required for simulation.  We consider both small systems which may be simulatable on near-future quantum devices, and systems sufficiently large for classical simulation to be intractable.  We use 86 molecular systems to demonstrate that the use of the Bravyi-Kitaev transformation is typically at least approximately as efficient as the canonical Jordan-Wigner transformation, and results in substantially reduced gate count estimates when performing limited circuit optimisations.
		
	\end{abstract}
	\section{Introduction}

	Computational chemistry is the use of well-developed theoretical techniques and algorithms to solve chemical problems.  These typically relate to the properties of molecules and chemical reactions.  Such processes occur as a result of the rearrangement of electrons among atoms.  Quantum chemistry is the branch of computational chemistry concerned with the theoretical understanding of these processes.~\cite{szabo_modern_1996}
	
	Although a vast spectrum of methods has been developed for this purpose, the field is restricted by the computational difficulty of the task.  Many calculations of interest involve the determination of ground state electronic wavefunctions and their corresponding energies.  To achieve this exactly (to within non-relativistic assumptions and basis set limitations) requires the \emph{full configuration interaction} approach.  This scales factorially with respect to the number of basis functions considered, limiting application of the technique to small molecules.~\cite{szabo_configuration_1996,gan_lowest_2006}

As this conceptually simple approach is computationally intractable, the difficulty of the task is often reduced by invoking various approximations, well-studied in computational chemistry.  While these methods often allow high degrees of precision, some computational tasks - for instance, in determining reaction kinetics or dynamics - would benefit from the decreased error of a full configuration interaction approach.
	
	Quantum simulation algorithms are expected to be capable of alleviating some of the difficulty associated with this through the use of a scalable quantum computer.  A quantum computer operates on \emph{qubits} -- the quantum equivalent of classical bits.  Instead of a unit which can either have a state value of $0$ or $1$, qubits exist as superpositions of $\ket{0}$ and $\ket{1}$ states, i.e. $\ket{\psi} = \alpha \ket{0} + \beta\ket{1}$.  A system of $n$ qubits can exist in a superposition of $2^n$ basis states.  Similar to classical computation, operations which manipulate the state of qubits are described as quantum gates and are analogous to classical logic gates.  A sequence of quantum gates, intended to perform a computational task, is referred to as a quantum circuit.  Gates that perform an operation which entangle the state of two or more qubits are called entangling gates.

		  Quantum algorithms to address various chemical tasks have been developed, including the determination of energy spectra~\cite{aspuru-guzik_simulated_2005}, reaction rates~\cite{lidar_calculating_1999,kassal_quantum_2009,reiher_elucidating_2017}, and reaction dynamics~\cite{kassal_polynomial-time_2008}.  Quantum simulation of quantum systems -- particularly chemistry -- is often cited as being one of the most significant potential uses of quantum computation.~\cite{olson_quantum_2017}
		  
		    The development of a scalable quantum computer is an extremely difficult task.  Demonstrations of the quantum simulation of electronic structure problems have mostly thus far been limited to the consideration of small hydrides using a minimal basis set.  These have been reported in photonic,~\cite{lanyon_towards_2010} NMR~\cite{du_nmr_2010} and superconducting~\cite{omalley_scalable_2016} devices.  The first fully scalable demonstration of this kind was performed in 2016.~\cite{omalley_scalable_2016}  Recently, the use of the variational quantum eigensolver algorithm~\cite{mcclean_theory_2016} has extended this to the simulation of beryllium hydride.~\cite{kandala_hardware-efficient_2017}
		    
		    However, recent hardware developments have yielded devices that are rapidly increasing in size.~\cite{omalley_scalable_2016,sete_functional_2016}  Devices of up to 50 qubits are likely to be available in the near future.~\cite{mohseni_commercialize_2017}  It is likely that such devices will bring the field close to the ability to perform clasically intractable chemistry calculations.~\cite{reiher_elucidating_2017}  For this, advances in both hardware and circuit design are necessary.

	The canonical quantum algorithm for the solution of the electronic structure problem involves several steps.~\cite{aspuru-guzik_simulated_2005}  Firstly, the molecular orbitals forming a basis for the electronic states must be represented using states of qubits.  The electronic Hamiltonian must then be mapped to an operator on the qubit Hilbert space.~\cite{jordan_uber_1928,bravyi_fermionic_2002,seeley_bravyi-kitaev_2012-5,tranter_bravyi-kitaev_2015}  Following this, a Trotter-Suzuki approximation~\cite{trotter_product_1959-3,suzuki_generalized_1976,hatano_finding_2005} is invoked to form an evolution operator which is implementable on the quantum device.  Finally, a phase estimation algorithm is invoked in order to ascertain the ground state molecular energy.~\cite{kitaev_quantum_1995}
	
		  Many algorithmic developments have been made to further this goal.  In particular, hybrid quantum-classical schemes have been shown to yield accurate results for a fraction of the cost of canonical quantum simulation techniques~\cite{mcclean_theory_2016,mcclean_exploiting_2014,omalley_scalable_2016,kandala_hardware-efficient_2017}.  However, these techniques still require a transformation of electronic states and operators to states of and operations upon qubits.  The two most commonly used forms of this transformation are the Jordan-Wigner transformation and the Bravyi-Kitaev transformation,~\cite{jordan_uber_1928,aspuru-guzik_simulated_2005,bravyi_fermionic_2002,seeley_bravyi-kitaev_2012-5,tranter_bravyi-kitaev_2015} although other constructions have been examined.~\cite{havlicek_operator_2017,setia_bravyi-kitaev_2017}
	
	In the asymptotic limit and without further circuit optimisation, the Bravyi-Kitaev transformation is known to have logarithmically superior scaling with respect to circuit length.~\cite{bravyi_fermionic_2002,seeley_bravyi-kitaev_2012-5}  An examination of the performance of this process requires generating descriptions of quantum circuits which would perform the simulation.  Initial assessments of this technique demonstrated a saving associated with the simulation of the hydrogen molecule in a minimal basis, a smallest-case example.~\cite{seeley_bravyi-kitaev_2012-5}  This saving was thus expected to also be present for larger chemical examples.  However, further investigation of methane revealed that overall gate savings were relatively modest, although substantial savings were yielded in terms of entangling gates.~\cite{tranter_bravyi-kitaev_2015}  To date, no large scale numerical analysis involving many systems has been performed. 
	
	In this paper, we use 86 molecular systems to demonstrate that the use of the Bravyi-Kitaev transformation typically results in quantum circuit lengths equal to or shorter than circuits using the canonical Jordan-Wigner transformation.  We also consider the impact of Trotter ordering upon both overall gate count and the relative performance of the Bravyi-Kitaev transformation.  Varying the Trotter ordering can impact the error incurred in the use of this approximation, potentially resulting in increased overall gate count, even if within each Trotter step the gate count is reduced.  As such, we consider the impact of Trotter ordering on the Trotter error, by examining a subset of our systems.

We begin by providing a brief overview of the theory underlying the Bravyi-Kitaev mapping and of Trotterization.  In section ~\ref{section:basic}, we discuss circuits with a limited degree of optimization.  Following this, we introduce the impact of Trotter ordering, considering its impact on single Trotter step circuit length in section \ref{section:trotterOrdering}, and on the Trotter error in section \ref{section:trotterError}.

	\color{black}
	\section{Theory}
	\color{black}
	The electronic Hamiltonian in the second quantized formalism is given by:
	
		\begin{equation}
		\hat{H} = \sum_{i,j}{h_{ij}a^\dagger_i a_j}+\frac{1}{2} \sum_{i,j,k,l}{h_{ijkl} a^\dagger_i a^\dagger_j a_k a_l}
		\label{eq:elecham}
		\end{equation}
		
	where $h_{ij}$ and $h_{ihkl}$ are Coulombic overlap and exchange integrals determined by the basis set chosen.~\cite{helgaker_molecular_2000,szabo_modern_1996}
	
	Although the number of $h_{ij}$ and $h_{ijkl}$ integrals scales quartically with respect to the number of basis orbitals, they are efficiently computable using conventional, classical computing methods.  Additionally, despite this theoretical quartic scaling, the number of nonzero integrals is often substantially reduced through consideration of molecular symmetries.~\cite{helgaker_molecular_2000,szabo_modern_1996}  Rather, the difficulty is due to the dimension of the Fock space on which this Hamiltonian acts, which scales exponentially with the number of basis orbitals considered.  Restricting the problem to a subspace with a fixed number of electrons reduces this scaling to being factorial with respect to the number of electrons, but for practical problems this remains intractable.  This difficulty typically prevents the use of full configuration interaction calculations for purposes other than benchmarking.
	
\begin{figure}[h!]
		\centering
		\includegraphics[width=\aetfigsize]{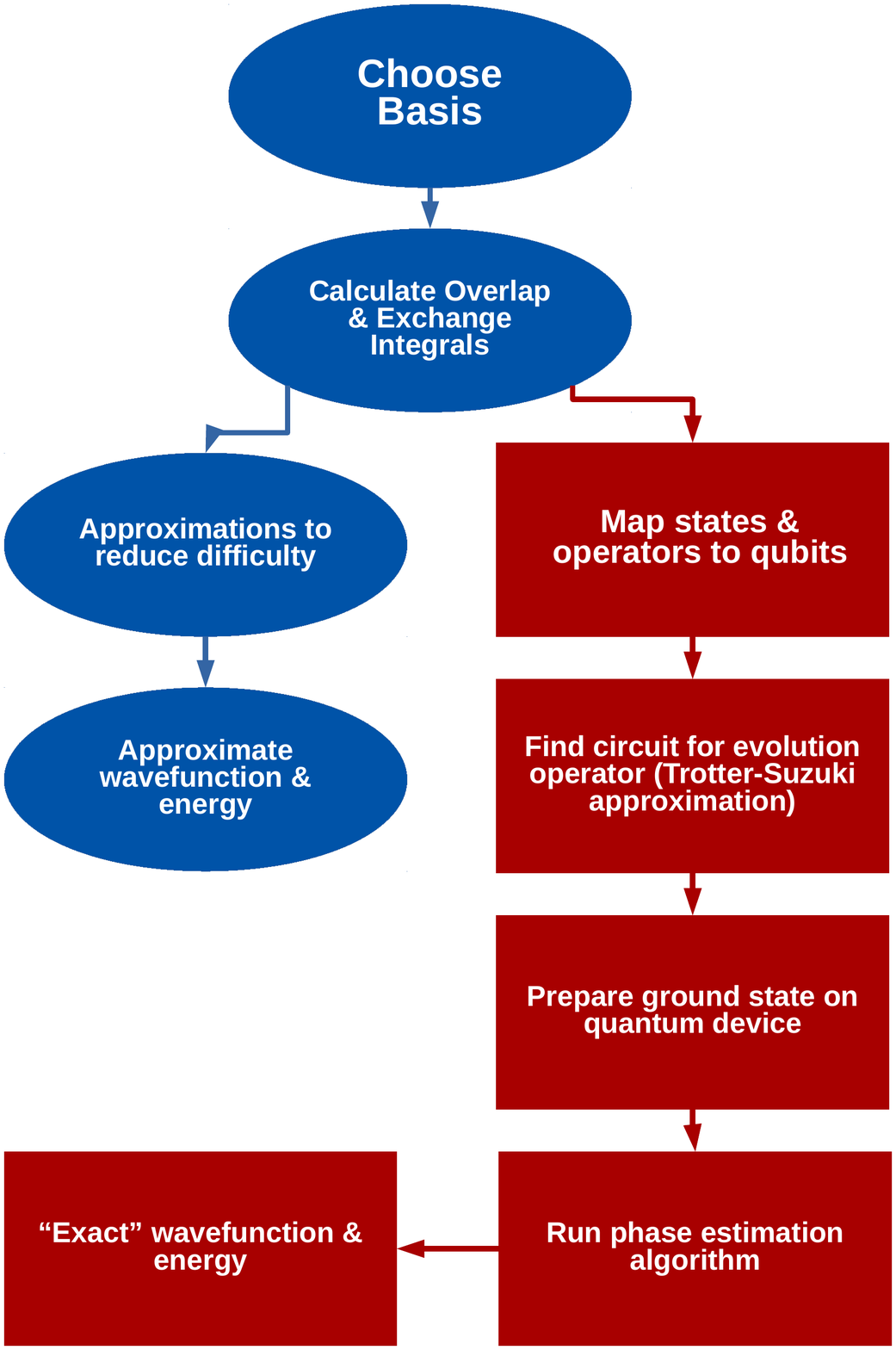}
		\caption{A comparison of classical and quantum algorithms for simulation of electronic structure.  Left: classical, Right: quantum}
		\label{fig:flowchart}
	\end{figure}

	On a quantum computer, this scaling difficulty is not present.  The task proceeds in four stages, illustrated in Figure ~\ref{fig:flowchart}.  First, a representation of the Hamiltonian and the molecular orbital spaces upon which it acts must be represented using states of and operations upon qubits.  Following this, a Trotter-Suzuki approximation~\cite{trotter_product_1959-3,suzuki_generalized_1976,hatano_finding_2005,somma_quantum_2003} is invoked in order to find a quantum circuit to simulate the evolution of the system under the molecular Hamiltonian.  The use of this approximation results in an additional error in simulation.  However, this error can be arbitrarily reduced through increasing the number of Trotter steps, resulting in only an extra multiplicative factor in the quantum computational expense.  A full consideration of Trotterization error is included in section~\ref{section:trotterError}. Having developed such a circuit, a good approximation to the simulated ground state of the system is prepared, and the phase estimation algorithm is used to ascertain the ground state energy of the system.  
	
We initially concern ourselves with the first of these stages - the mapping technique chosen to transform electronic states and operators to states and operators of qubits.  We discuss the difference in resource implications for two options for this:  the canonical Jordan-Wigner transformation, and the Bravyi-Kitaev transformation.  In section~\ref{section:trotterError} we address the implications of these mappings when performing a Trotter-Suzuki approximation.
	
	Our task in establishing an appropriate mapping is to find qubit analogues of both the electronic states and the creation and annihilation operators in eqn~(\ref{eq:elecham}). Traditionally, the simplest encoding scheme to determine these is the \emph{Jordan-Wigner transformation}.~\cite{jordan_uber_1928,aspuru-guzik_simulated_2005}  Here, $n$ qubits are used to store the occupation number of $n$ electronic spin-orbitals, forming what is known as the occupation basis.  If the $i$th molecular orbital is occupied, then the corresponding $i$th qubit is in the $\ket{1}$ state, whereas if the molecular orbital is unoccupied, then the qubit is in the $\ket{0}$ state.  We then require a representation of the electronic creation and annihilation operators that act on the qubit space, which perform the following set of operations:
	
		\begin{equation}
		\begin{aligned}[c]
		\hat{Q}^+\ket{0}&=\ket{1}\\
		\hat{Q}^-\ket{0}&=0
		\end{aligned}
		\qquad\qquad\qquad
		\begin{aligned}[c]
		\hat{Q}^+\ket{1}&=0\\
		\hat{Q}^-\ket{1}&=\ket{0}
		\end{aligned}
		\end{equation}
	
	A naive assessment would suggest that the standard Pauli $\sigma^+_i$ and $\sigma^-_i$ operators would suffice for this purpose; however, these do not obey the required anticommutation relations:
	\begin{eqnarray}
	\{a^\dagger_i,a^\dagger_j\} = \{a_i,a_j\} = 0 \\
	\{a^\dagger_i,a_j\} = \delta_{ij}I 
	\end{eqnarray}
	
	For this these to hold, the parity of the occupation numbers of the orbitals with index less than $i$ must be calculated, and a phase shift introduced when the parity is odd.  This is accomplished by performing a sequence of Pauli $Z$ operations on the preceding qubits, yielding the following:
		\begin{eqnarray}
		a^\dagger_i = \frac{1}{2} \left( X_i - i Y_i \right) \bigotimes_{j<i}  Z_j \nonumber \\
		a_i = \frac{1}{2} \left( X_i + i Y_i \right) \bigotimes_{j<i}  Z_j
		\end{eqnarray}
	
	 Note that this mapping requires $\mathcal{O}(N)$ qubit operations to simulate one electronic operation.
	
	An alternative scheme was envisaged by Bravyi and Kitaev whereby parity information is stored in the qubit states - i.e. qubit $i$ stores the sum (modulo 2) of the occupation of all electronic states with index less than or equal to $i$.  This basis - the \emph{parity basis} - avoids the additional cost of determining the parity, as this information can be queried with only a single qubit operation.~\cite{bravyi_fermionic_2002,seeley_bravyi-kitaev_2012-5}  However, this mapping has instead delocalised information regarding the occupation of each electronic orbital.  Clearly, any electronic creation or annihilation operation on an orbital with index $i$ requires the update of all qubits with index greater than or equal to $i$.  Consequently, using this mapping the number of qubit operations required to simulate one electronic operation is also $\mathcal{O}(N)$.
	
	\subsection{The Bravyi-Kitaev Mapping}
	
	The Bravyi-Kitaev~\cite{bravyi_fermionic_2002,seeley_bravyi-kitaev_2012-5,tranter_bravyi-kitaev_2015}  mapping is an attempt to improve upon the linear scaling of the occupation and parity bases.  In essence, it is a middle ground between these approaches.  For a molecular orbital basis of size $N$, there are again $N$ qubits used.\bibnote{This requirement can be relaxed by additional encoding techniques,~\cite{bravyi_tapering_2017,steudtner_lowering_2017} however we do not consider these techniques here.}   However, the information stored within each qubit now varies dependent on the index $i$.  Note that we begin indexing at $i=0$.  Where $i$ is even, qubit $i$ stores the occupation number of orbital $i$, as in the Jordan-Wigner mapping.  Where $i$ is odd, the qubit stores the parity of a particular set of occupation numbers.  When $\log_2 \left(i+1\right)$ is an integer, the qubit stores the parity of the occupation numbers of all orbitals with indices less than or equal to $i$.  For other cases, the qubit stores the parity of the occupation numbers of orbitals in subdividing binary sets.  This complex mapping is best understood through consideration of the matrix which transforms a vector of orbital occupations to qubit states.  For example, in the eight orbital/qubit case, this is given by:
		\begin{equation}
		\begin{bmatrix}
		1 & 0 & 0 & 0 & 0 & 0 & 0 & 0 \\
1 & 1 & 0 & 0 & 0 & 0 & 0 & 0 \\
0 & 0 & 1 & 0 & 0 & 0 & 0 & 0 \\
1 & 1 & 1 & 1 & 0 & 0 & 0 & 0 \\
0 & 0 & 0 & 0 & 1 & 0 & 0 & 0 \\
0 & 0 & 0 & 0 & 1 & 1 & 0 & 0 \\
0 & 0 & 0 & 0 & 0 & 0 & 1 & 0 \\
1 & 1 & 1 & 1 & 1 & 1 & 1 & 1 \\
		\end{bmatrix}
		\begin{bmatrix}
		o_0 \\
		o_1 \\
		o_2 \\
		o_3 \\
		o_4 \\
		o_5 \\
		o_6 \\
		o_7
		\end{bmatrix}
		= 
		\begin{bmatrix}
		o_0 \\
		o_0 + o_1 \\
		o_2 \\
		o_0 + o_1 + o_2 + o_3 \\
		o_4 \\
		o_4 + o_5 \\
		o_6 \\
		o_0 + o_1 + o_2 + o_3 + o_4 + o_5 + o_6 + o_7 
		\end{bmatrix}
		=
		\begin{bmatrix}
		q_0 \\
		q_1 \\
		q_2 \\
		q_3 \\
	    q_4 \\
		q_5 \\
	    q_6 \\
		q_7
		\end{bmatrix}
		\label{eqn:bk}
		\end{equation}
	Here, each $o_i$ value corresponds to the occupation number of the orbital with index $i$, and the $q_i$ values correspond to the state of the qubit with index $i$.  Where $q_i$ is 0, qubit $i$ is in the $\ket{0}$ state and similarly where $q_i$ is 1, qubit $i$ is in the $\ket{1}$ state.  All sums are performed modulo 2.  The matrix on the left is thus the matrix which transforms orbital occupation numbers to qubit states.

	Both occupation and parity information is now stored non-locally.  Inspection of eqn~(\ref{eqn:bk}) shows that this information is stored in a number of qubits which grows logarithmically.  Thus, any electronic creation or annihilation operation can be simulated in $\mathcal{O}(\log_2{n})$ qubit operations.  We omit a detailed proof of this here for reasons of brevity.  Further details can be found in references.~\cite{bravyi_fermionic_2002,seeley_bravyi-kitaev_2012-5,tranter_bravyi-kitaev_2015}
	
	Despite the superior asymptotic scaling of the Bravyi-Kitaev mapping, it is important to consider the increased overhead of its use.  Initial implementations noted that the Bravyi-Kitaev mapping is more efficient than the Jordan-Wigner mapping in the simulation of molecular hydrogen in a minimal basis, the smallest possible non-trivial example.~\cite{seeley_bravyi-kitaev_2012-5}  It was thus argued that the overhead was not a significant factor, and the superior scaling effectively dominated in all cases.  However, further investigation on methane in a minimal basis revealed that this is not the case.~\cite{tranter_bravyi-kitaev_2015}  
	
	One purpose of this paper is to find the point at which this asymptotically superior scaling dominates.  
	Examination of the Bravyi-Kitaev creation and annihilation operators permits a rough estimate of this.  Note that the qubit creation and annihilation operator equivalents using the Bravyi-Kitaev transformation are given by:
	
\begin{eqnarray}
a^\dagger_i =
\frac{1}{2}
 \left(  X_{U(i)} \otimes X_i \otimes Z_{P(i)} - i X_{U(i)} \otimes X_i \otimes Z_{P(i)} \right)\\
 a_i =
 \frac{1}{2}
 \left(  X_{U(i)} \otimes X_i \otimes Z_{P(i)} + i X_{U(i)} \otimes X_i \otimes Z_{P(i)} \right)
\end{eqnarray}

	where $U(i)$ is the ``update set" of qubit $i$, and $P(i)$ is the ``parity set" of qubit $i$.  For brevity we do not discuss these sets here, however their size is of maximum order $\mathcal{O}(\log_2{i})$.  These expressions are valid only in the case of even $i$.  However, this does not affect our rough estimate.  Examining these equations, it is evident that at most $4\log_2{i} + 2$ gates are required for the simulation of one fermionic operation.  This quantity is smaller than the simple $i$ gates of the Jordan-Wigner mapping when $i \geq 19$.  Thus, noting that the Bravyi-Kitaev mapping is most efficient when $N$ approaches a power of $2$ (as it can take increased advantage of its binary tree structure), we conservatively estimate that this point should be at $N\approx32$.  We thus would expect quantum computational cost to be reduced when using the Bravyi-Kitaev transformation for systems with more than $N\approx32$ spin-orbitals.
	
\color{black}
	\subsection{Trotterization and Simulation}
	
	In order to perform the phase estimation algorithm to determine the molecular ground state energy, a quantum circuit simulating the unitary evolution operator $\hat{U} = \exp{\left(-it\hat{H}/\hbar\right)}$ of the molecular Hamiltonian must be found.  This is similarly required when utilising a variational quantum eigensolver algorithm using a coupled-cluster ansatz.~\cite{peruzzo_variational_2014,mcclean_theory_2016,omalley_scalable_2016,kandala_hardware-efficient_2017}  The qubit form of the electronic Hamiltonian determined through the Bravyi-Kitaev or Jordan-Wigner transformation consists of a weighted sum of strings of Pauli operations.  In order to exponentiate this, a Trotter-Suzuki approximation must be invoked.~\cite{suzuki_generalized_1976}  The first order Trotter-Suzuki expansion is:
	
	\begin{equation}
	e^{\frac{-it}{\hbar}\sum_i{\hat{H}_i}}  \approx \left( \prod_i{e^{\frac{-it\hat{H}_i}{\hbar n}}} \right)^n
	\label{eq:trotter}
	\end{equation}

	where $H_i$ are the Hamiltonian sub-terms (strings of Pauli operations, in our case) and $n$ is the number of Trotter steps.  The overall evolution time is now subdivided into $n$ steps.  Increasing the number of Trotter steps decreases the error invoked in this procedure.
	\begin{figure}[h]
	\centering
	\includegraphics[width=\aetfigsize]{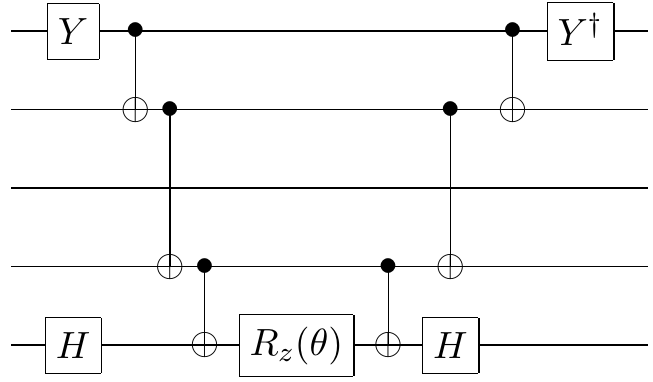}
	\caption{The canonical circuit for the simulation of $exp(-i\frac{\theta}{2}Y_0Z_1Z_3X_4)$}
	\label{plot:circ1}
\end{figure}	
	This yields the evolution operator expressed as a product of exponentiated strings of Pauli operations.  Standard techniques can be used to transform these into quantum circuits, as shown in Figure ~\ref{plot:circ1}.  The gates within this circuit can be divided into two types:  gates that rotate the state of a single qubit, and typically more expensive two-qubit entangling gates.  These can be implemented sequentially to form a quantum circuit which simulates the entire evolution operator.
	
	The use of a Trotter-Suzuki approximation results in the introduction of error.~\cite{poulin_trotter_2015,babbush_chemical_2015,whitfield_simulation_2011}  This error can be reduced by increasing the number of Trotter steps considered.  We consider the impact of this error in section ~\ref{section:trotterError}.

	\section{Basic Circuits}
	\label{section:basic}
	Our code was used to assess the serial quantum circuit length  corresponding to the simulation of 86 small molecules and atoms.  Molecular structures were gathered from the NIST CCCBDB database~\cite{johnson_iii_nist_2016} optimised at the Hartree-Fock level.  Most systems used a STO-3G basis; however, larger Pople basis sets were used in 18 trials.  Of these, four systems (\ce{CH2^{2.}}, \ce{HF}, \ce{LiH}, \ce{H2O}) using a 3-21G basis set were examined, with the remainder studying \ce{H2} and \ce{HeH+}.  Clearly, this choice of basis is insufficient for an accurate solution of the electronic structure problem.  When performing a simulation on a real quantum device, a larger basis set would be chosen as in conventional quantum chemistry methods.  Fortunately, the error introduced by our choice of basis is fixed and independent of our choice of quantum methodology.  As our benchmarking procedure is not directly concerned with the exact energies predicted, our choice of basis set allows for the simulation of a variety of systems with relatively low computational overhead.  Details of the systems studied can be found in Table ~\ref{table:systems1} and in the appendices.  Our systems range in size from $2$ to $54$ spin-orbitals.  While containing systems that are classically intractable to simulate, this number is relatively modest in contrast to simulations that would be performed upon a real quantum device~\cite{reiher_elucidating_2017}; however, it allows us to maintain relatively low computational expense (approximately one week of CPU time for the largest examples).
	
		\csvstyle{miniSystemsTableStyle}{
		late after line=\\,
		late after last line=\\\bottomrule,
		head to column names}
	\begin{table}
	\begin{tabular}{lccccccc}
	\toprule\bfseries Qubits & \bfseries 1-10 & \bfseries 11-20 & \bfseries 21-30 & \bfseries 31-40 & \bfseries 41-50&\bfseries 51-60 & \bfseries Total \\\midrule
	\begin{filecontents*}{systems.csv}
AETnumq,AETqa,AETqb,AETqc,AETqd,AETqe,AETqf,AETtotal
Molecules + Radicals,1 (1),13 (13),12 (3),4 (0),8 (0),3 (0),41 (17)
Atoms,10 (4),7 (6),1 (0),1 (0),0 (0),1 (0),20 (10)
Ions,2 (2),4 (3),0 (0),2 (0),1 (0),1 (0),10 (5)
Non-minimal basis sets,4 (4),6 (5),5 (2),0 (0),0 (0),0 (0),15 (11)
,,,,,,,
Total,17 (11),30 (27),18 (5),7 (0),9 (0),5 (0),86 (43)
\end{filecontents*}
\csvreader[miniSystemsTableStyle]
	{systems.csv}{}{\AETnumq & $\AETqa$ & $\AETqb$ & $\AETqc$ & $\AETqd$ & $\AETqe$ & $\AETqf$ &$\AETtotal$}%
	\end{tabular}
	\captionsetup{width=.9\linewidth}
	\caption{A breakdown of systems studied.  Note that most of the systems involving a non-minimal basis set were H\textsubscript{2} and HeH\textsuperscript{+} systems, as specified in the appendices. Numbers in parenthesis indicate the number of systems where Trotter error was considered, as discussed in section ~\ref{section:trotterError}.}
	\label{table:systems1}
	\end{table}

	Hartree-Fock molecular orbitals and their $h_{ij}$ and $h_{ijkl}$ integrals were obtained using the PSI4 electronic structure theory package~\cite{parrish_psi4_2017} and the FermiLib PSI4 Plugin~\cite{mcclean_openfermion_2017}.  Our code was then used to generate Jordan-Wigner and Bravyi-Kitaev Hamiltonians.  These are stored symbolically as strings of Pauli operations, as in previous work.~\cite{seeley_bravyi-kitaev_2012-5,tranter_bravyi-kitaev_2015}  The Hamiltonian can be stored as blocks of second quantised terms, potentially grouped according to their character - excitation operations, number operations, and so on.  Note that due to molecular symmetries (and the symmetries of the integrals present in eqn.~\ref{eq:elecham}), the terms in eqn.~\ref{eq:elecham} do not necessarily have independent coefficients.  
	
	A basis of Hartree-Fock molecular orbitals was used to describe the system when employing the Jordan-Wigner or Bravyi-Kitaev mapping.  Much work has been performed in assessing the performance of other basis choices~\cite{mcclean_exploiting_2014,babbush_chemical_2015}, with localised basis orbitals showing promise in reducing the number of significant terms in the Hamiltonian.  However, as this advantage is gained from reduction of the number of significant overlap integrals, there is no obvious reason to believe that the Jordan-Wigner and Bravyi-Kitaev mappings would perform inequivalently in a predictable manner.  Preliminary analysis using natural and orthogonalised atomic orbital integrals provided by collaborators~\cite{babbush_chemical_2015} did not suggest any consistent dependence of the performance of the Bravyi-Kitaev mapping (versus the Jordan-Wigner mapping) on the choice of basis considered.
	
As such, in order to reduce the computational cost of our simulations, this degree of freedom was not considered here.  A rigorous demonstration of the independence of the performance of the Bravyi-Kitaev mapping on the basis choice could be considered in future work.
	
	\color{black}
	Optimisation and Trotterization could be performed on the level of second quantized operators.  Other works have taken this approach, maintaining fermionic terms throughout optimisation procedures.~\cite{poulin_trotter_2015,hastings_improving_2015}  Instead, our code does not retain the original fermionic components of the electronic Hamiltonian, and reduces the qubit Hamiltonian to a list of strings of Pauli operations, before attempting circuit-level optimisation.  While fewer assumptions can be made about the structure of the new Hamiltonian (a fact of relevance in section~\ref{section:further}), this approach allows for greater flexibility when ordering terms for Trotterization.
	
	From here, our code allows for generation of quantum circuit objects corresponding to the implementation of one Trotter step of the evolution operator of the qubit Hamiltonian (discussed below).
	
	Neglecting any benefits from optimisation at the interfaces of Trotter steps (which would save at most $\mathcal{O}(N_{trotter})$ gates), the number of gates necessary for larger numbers of Trotter steps is a simple multiple of the number of gates necessary for one Trotter step.  Consequentially, extending our analysis to higher Trotter numbers was not considered necessary for our initial analysis, although this was performed when considering Trotter error in section~\ref{section:trotterError}.
	
  A full treatment of the entire phase estimation algorithm (including error correction) was not performed.  This is in contrast to other work that has attempted to characterise the resource implications of performing the full procedure.~\cite{reiher_elucidating_2017,poulin_trotter_2015}  For our analysis of the raw gate counts of Bravyi-Kitaev circuits, this was not considered important for the above reasons.  Note that at present, the largest simulations conducted have been of around 45 qubits.~\cite{haner_0.5_2017,pednault_breaking_2017,smelyanskiy_qhipster_2016}  These required extensive specialised computing resources.  Performing simulations at this level would have made it impossible to perform a large, multi-system survey.  In particular, assessing the point at which Bravyi-Kitaev scaling is expected to dominate (around 32 qubits, as discussed above) would have been problematic.  Despite this, a full consideration of the phase estimation algorithm could yield useful results in regard to the Trotter error of simulation.  For simplicity, we have only considered here the canonical phase estimation procedure, as the expected benefits of the Bravyi-Kitaev mapping will apply in any system which involved the use of such a mapping scheme.
 
   It is important to note that many of the circuits discussed in this paper are substantial in terms of quantum resources required.  For around $50$ spin-orbitals (and thus $50$ logical qubits), the unoptimised circuits consist of around $10^7$ gates.  It is likely that the implementation of such circuits on a quantum device would require the use of some form of error correcting code.  In order to assess resource implications of circuits within a fault tolerant framework, the number of Clifford and non-Clifford gates within the circuit are counted.~\cite{nielsen_quantum_2010}  While Clifford gates are considered relatively straightforward to implement in a fault-tolerant manner, the resource implications of performing the non-Clifford gates are assessed by counting the number of T ($\pi/4$ phase rotation) gates required to implement them.
   
While a thorough analysis of the practicality of implementing these circuits on a quantum device would require consideration of this point, our focus here is on assessing the use of the Bravyi-Kitaev mapping.  Observing Figure~\ref{plot:circ1}, it is evident that only the central rotation gate is a non-Clifford gate.  There is one of these gates for every term in the qubit Hamiltonian.  As the number of terms in the Hamiltonian is the same for either mapping scheme, this implies that the number (and type) of non-Clifford gates is the same regardless of the mapping scheme chosen.  This is confirmed by numerical analysis in the circuits discussed below.  This implies that the choice of Bravyi-Kitaev or Jordan-Wigner mapping does not impact that T count of the circuit, and thus does not impact the difficulty of performing error correction.  As such, we do not consider this difficulty here.

 However, previous studies have shown~\cite{seeley_bravyi-kitaev_2012-5,tranter_bravyi-kitaev_2015} that the Bravyi-Kitaev mapping results in a reduction in the number of CNOT gates required, independent of those used in constructing a fault-tolerant representation of the central rotation gate.  This comes at the expense of an increased number of single qubit Clifford gates required.  In other words, these previous examples showed that the Bravyi-Kitaev mapping traded a reduction in two qubit Clifford gates for an increase in one qubit Clifford gates (along with an overall decrease in the total number of Clifford gates).  While this is not of huge impact in a fault tolerant framework (as the T count remains the same), experimental devices in the near future are still likely to benefit from the minimization of entangling gates so as to reduce error.  As such, we have considered the breakdown of circuits into entangling and single qubit gates in this paper.
 
\color{black}
 
\begin{figure}[h!]
	\centering
	\includegraphics[width=\aetfigsize]{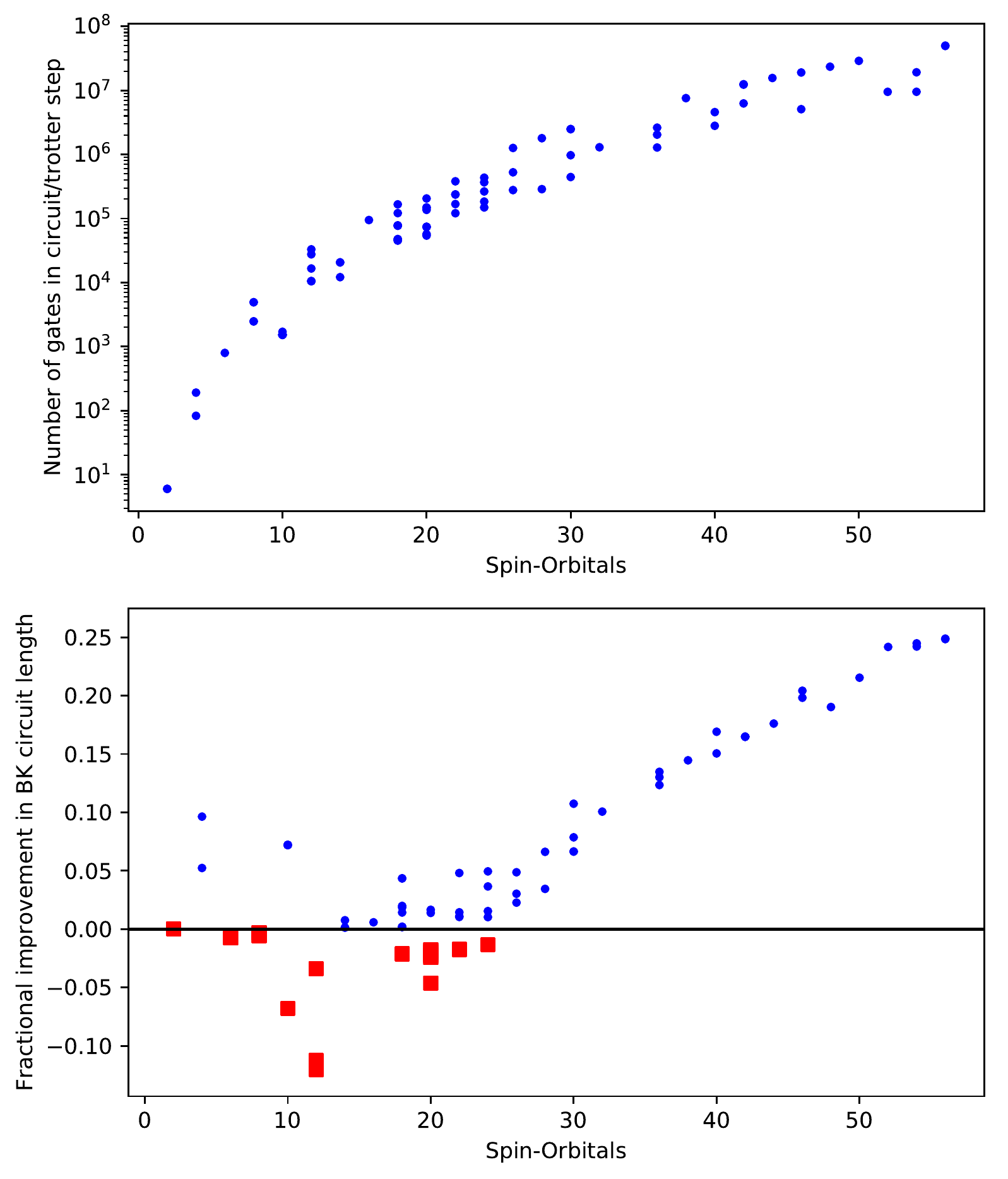}
	\caption{Unoptimised gate counts.  Upper: Total number of gates in Jordan-Wigner circuits, before optimisation.  Lower: Gate savings through use of Bravyi-Kitaev mapping as a fraction of Jordan-Wigner gate count.  Squares indicate instances where the Jordan-Wigner transformation outperformed the Bravyi-Kitaev transformation.  In this scheme, the Bravyi-Kitaev transformation reliably results in shorter circuits from around 30 spin-orbitals, with up to around 25\% gate savings in the examples with 50 spin-orbitals.}
	\label{plot:initialGates}
\end{figure}	

	Initially, the Hamiltonians were ordered by the magnitude of their coefficients.  This somewhat arbitrary ordering was chosen in order to assess the pre-optimisation efficacy of the Bravyi-Kitaev mapping, and is in contrast to optimised ordering schemes used in other work.~\cite{hastings_improving_2015,seeley_bravyi-kitaev_2012-5} These are considered in detail in sections 4 and 5.  The systems studied involved between $2$ (the hydrogen atom) and $56$ (the iodine atom) spin-orbitals.  As to be expected,~\cite{poulin_trotter_2015} the serial circuit length dramatically increases for larger systems, requiring of order $10^7$ gates for the simulation of systems involving around $50$ spin-orbitals.  While not as ruinous as the factorial difficulty of classical full configuration interaction, this large circuit length illustrates the need for compiler optimisations.

	An initial assessment of circuits for the implementation of Jordan-Wigner and Bravyi-Kitaev Hamiltonians suggests that the use of the Bravyi-Kitaev mapping is associated with a substantial improvement for the larger systems.  Figure ~\ref{plot:initialGates} demonstrates this.  From roughly 30 spin-orbitals, this improvement is consistent, and constitutes approximately 25\% of the overall circuit length for the largest of the systems we have examined. However, many systems smaller than this see no improvement, or even demonstrate larger circuit lengths.  This is in line with our earlier rough estimate that the superior scaling of the Bravyi-Kitaev mapping dominates the increased overhead at around 32 spin-orbitals.  Classical full configuration interaction calculations have been performed on systems marginally larger than this (36 molecular orbitals).~\cite{rossi_full-configuration_1999}  Consequently, for simulations aiming to achieve results which are classically intractable, a naive approach involving no circuit optimisations would be substantially eased through the use of the Bravyi-Kitaev mapping.

	In addition to optimisation performed through combination of duplicate Pauli strings, our code optimises circuits by the cancellation of duplicate gates.~\cite{hastings_improving_2015}  Circuit objects can automatically search their gate sequence for duplicate self-inverse gates and remove them.  Furthermore, the code tests individual gates for commutativity with gates that follow in sequence.  If such commutativity is present, it tests to see if any accessible gates are accessible through commutation.  This is performed according to a set of rules: gates acting on different qubits always commute, CNOT gates commute unless one targets the other's control, and so on.  This avoids the generation of matrix representations of the gates.  Optimisation in this form is repeated until the circuit is self-consistent and no further optimisation could be yielded.\bibnote{This iterative process is not optimised for classical computing cost, and would likely be performed more efficiently when generating circuits to be implemented on real quantum devices.}
	
	Based on work by Hastings, Wecker, Bauer and Troyer~\cite{hastings_improving_2015,poulin_trotter_2015}, savings from this procedure arise from two factors.  Firstly, redundancy in parity determination is eliminated, as this information is not decomputed after every term in the Hamiltonian.  This results in savings in the CNOT string used to determine parities.  Additionally, basis changes are not decomputed when unnecessary.  This saves on the single qubit $H$ and $Y$ gates necessary to set these bases.

\begin{figure}[H]
	\centering
	\includegraphics[width=\aetfigsize]{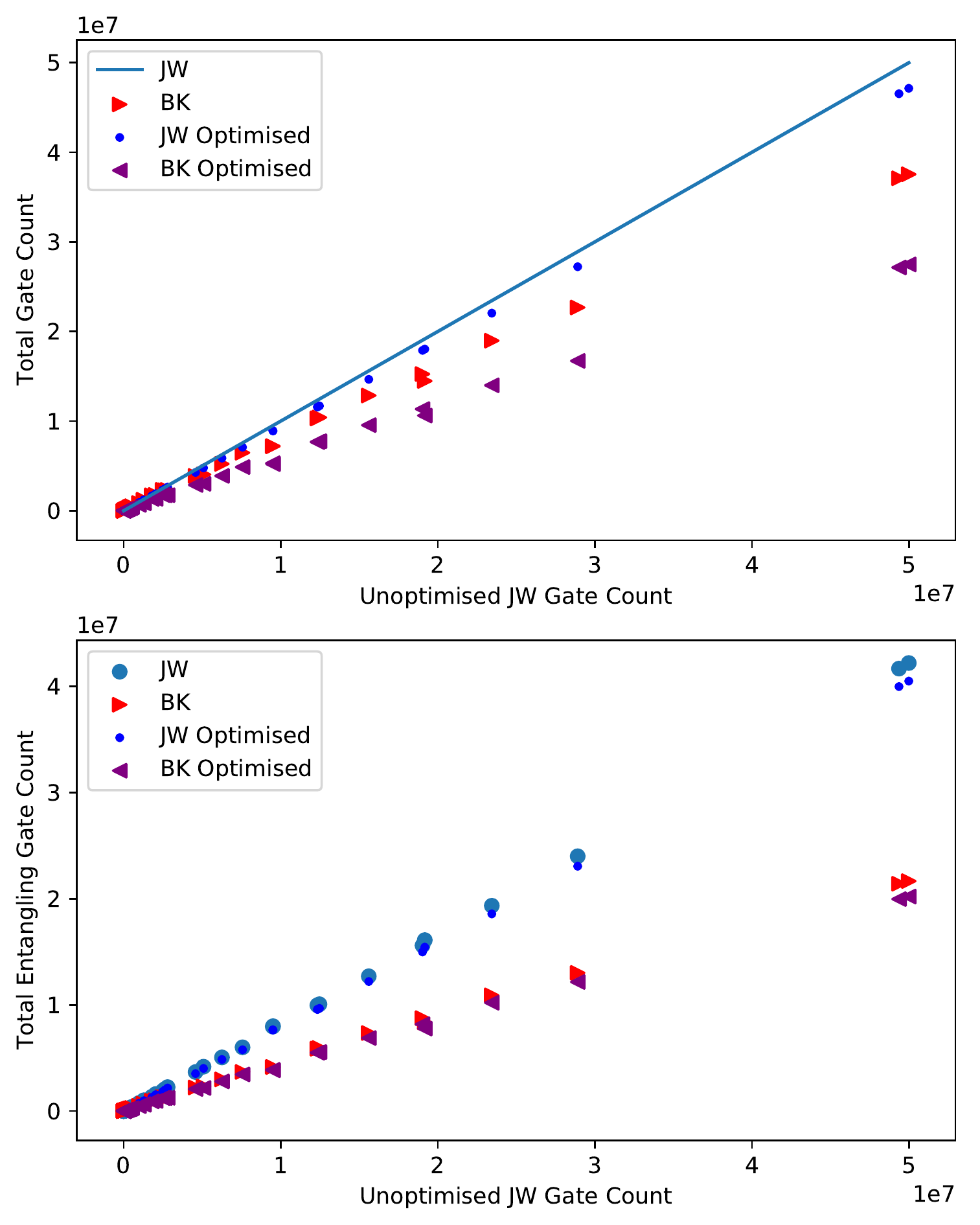}
	\caption{Number of gates in Bravyi-Kitaev and Jordan-Wigner circuits, before and after gate cancellation, versus the number of gates in Jordan-Wigner circuit before optimisation, using a magnitude ordering.  Upper: Total gate count.  Lower: Entangling gate count only.}
	\label{plot:standardGateComparison}
\end{figure}

When duplicate gates in the circuit are removed, the superiority of the Bravyi-Kitaev mapping is even more pronounced while still using a magnitude ordering.  This relative improvement appears to increase with larger circuits, as demonstrated by Figure~\ref{plot:standardGateComparison}.  In circuits involving more than about $10^7$ gates, the reduction in gate count associated with the use of the Bravyi-Kitaev mapping is larger than that of gate cancellation.  In these cases, the use of the Bravyi-Kitaev mapping results in circuits that are approximately 30-40\% shorter.  Additionally, the number of gates cancelled using the Bravyi-Kitaev mapping is several times greater than the number of gates cancelled using Jordan-Wigner mapping.  In some cases the advantage associated with the Bravyi-Kitaev mapping reduces the circuit length to that observed for systems involving fewer orbitals.  For example, the Bravyi-Kitaev circuit for the simulation of the iodine atom (54 spin-orbitals) requires 5204912 gates per Trotter step, whereas the Jordan-Wigner circuit for the simulation of acetone (52 spin-orbitals) requires 8954933 gates per Trotter step.

With systems requiring more than $10^6$ gates to simulate, there are no examples where the optimised Jordan-Wigner technique outperforms the optimised Bravyi-Kitaev technique.  Thus, using this magnitude ordering, it is clear that the Bravyi-Kitaev mapping should be preferred to the Jordan-Wigner mapping in the general case.

As discussed above, previous work~\cite{tranter_bravyi-kitaev_2015} on the methane molecule indicated that the Bravyi-Kitaev mapping may be advantageous with particular regard to the number of entangling gates required.  Our findings here confirm that this advantage holds in general, as shown in Figure ~\ref{plot:standardGateComparison}.  In addition to the general gate savings associated with the Bravyi-Kitaev mapping, we observe a substantial decrease in the number of CNOT gates required.  We consider this to be a major advantage of the Bravyi-Kitaev mapping.  This advantage is typically offset by a small increase in the number of single qubit gates required (as the total savings are smaller than the entangling gate savings).

While using a magnitude ordering, gate cancellation does not result in a great deal of entangling gate savings, which are typically far fewer than the advantage associated with the use of the Bravyi-Kitaev transformation.  Instead, the bulk of gate savings associated with cancellation techniques arises from maintaining the calculation basis between sequential terms, as opposed to continually resetting to the computational basis.  It is likely that many CNOT strings are being ``trapped" behind non-commuting gates earlier in their respective CNOT strings.  This could be alleviated by further circuit optimisation; however, this task is difficult to perform without further decreasing the locality of the CNOT chain.

The results of this scheme use a magnitude ordering for both the Jordan-Wigner mapping and the Bravyi-Kitaev mapping.  Further analysis of the performance of the Bravyi-Kitaev mapping is impossible without consideration of the Trotterization ordering chosen.  This is considered in the following section.  Manipulation of the Trotter ordering can also cause variation in the Trotter error, which could result in an increased number of Trotter steps necessary for constant precision.  This increased difficulty could undermine the savings gained from the use of a particular ordering in terms of cancellation, and is examined in section~\ref{section:trotterError}.

\section{Impact of Trotter ordering}
\label{section:trotterOrdering}
As discussed above, the overall goal is to find a minimal length circuit that can implement the unitary evolution operator of the quantum Hamiltonian.  As no standard circuit for the simulation of the exponentiated total Hamiltonian exists, a Trotter-Suzuki approximation must be invoked (eqn.~\ref{eq:trotter}).

The ordering of terms in this approximation is important.  It has been demonstrated~\cite{babbush_chemical_2015} that the error due to the utilisation of Trotter-Suzuki formulae strongly depends on the term ordering chosen.  Additionally, the number of duplicate gates depends strongly on the ordering chosen.  Both of these factors influence the length of the overall quantum circuit.

 The previous sections utilised a magnitude ordering of Trotter terms.  This ordering is significantly physically meaningful, as terms with higher magnitude are likely to correspond to stronger physical interactions.  However, it is also known to be suboptimal in terms of gate cancellation procedures.~\cite{hastings_improving_2015} 

As the number of potential orderings grows factorially, the problem of finding an optimal ordering scheme is a difficult one.  However, ordering schemes that are superior for the process of gate cancellation can be found, as the similarity of sequential Pauli strings  - and thus the savings from cancellation - can be determined when specifying the Hamiltonian.  Our analysis compares the impact of the use of the Bravyi-Kitaev mapping with a varying choice of ordering.

A lexicographic ordering is expected to maximize the gate savings obtained through cancellation, as the similarity of adjacent terms is maximized.~\cite{hastings_improving_2015}  As our code optimises on the level of Pauli strings rather than fermionic operators, we order on this level also, with no ordering performed on the fermionic level.  Our code stores Pauli strings as lists of base 4 integers.  A lexicographic ordering in this scheme is then essentially a bitwise numerical ordering.

\begin{figure}[H]
	\centering
	\includegraphics[width=\aetfigsize]{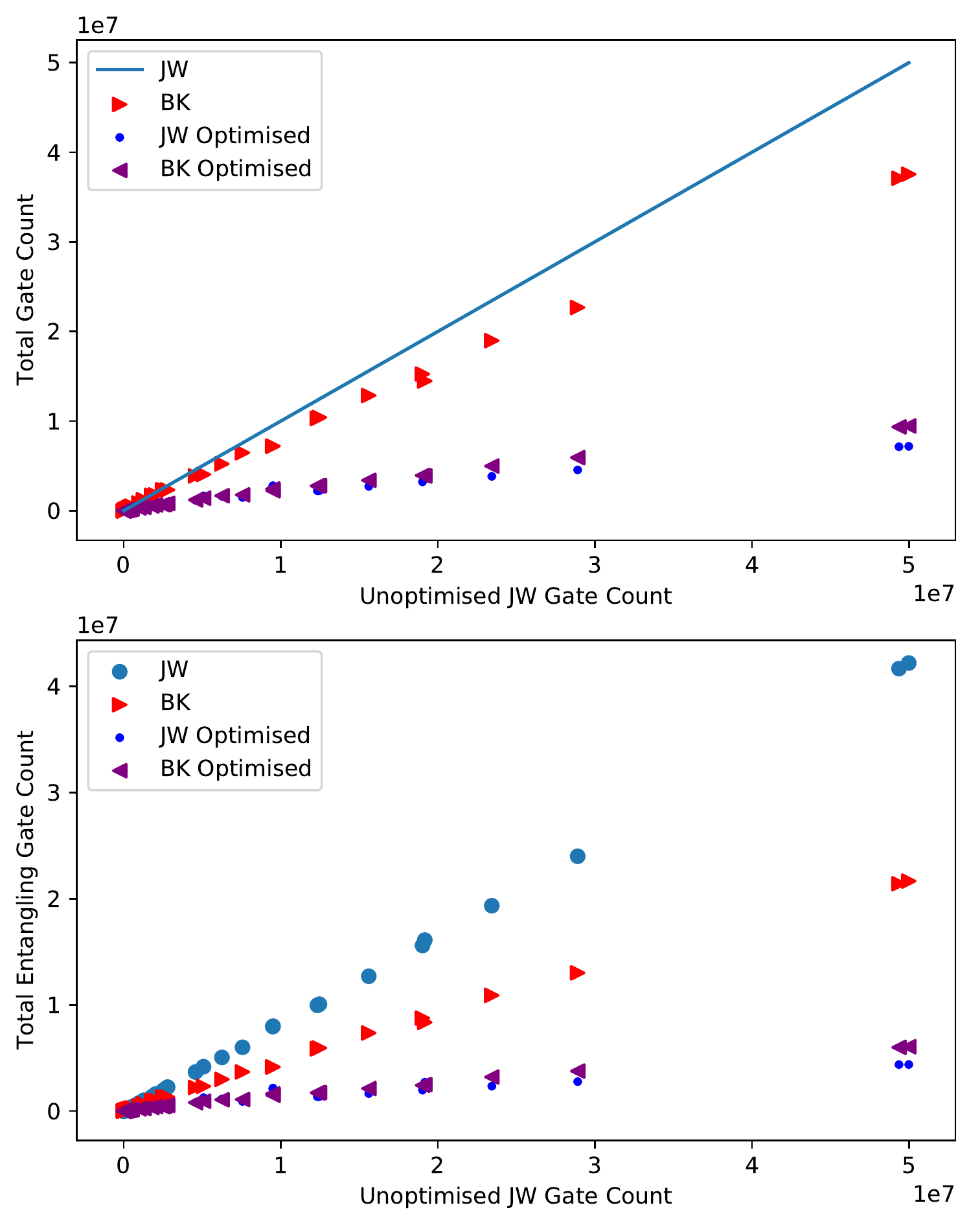}
	\caption{Number of gates in Bravyi-Kitaev and Jordan-Wigner circuits, before and after gate cancellation, versus the number of gates in Jordan-Wigner circuit before optimisation, using a lexicographic ordering.  Upper: Total gate count.  Lower: Entangling gate count only.}
	\label{plot:topLexGateComparison}
\end{figure}

We present results based on this ordering explicitly in Figure ~\ref{plot:topLexGateComparison}.  Note firstly the dramatic gate savings associated with using this optimisation and ordering scheme.  Whereas using a magnitude ordering, the Bravyi-Kitaev mapping provided the bulk of the gate savings once gate cancellation had been performed, now the impact of the Bravyi-Kitaev mapping is relatively minor.  The savings associated with the use of a lexicographic ordering far outweigh the savings associated with the of the Bravyi-Kitaev mapping with a magnitude ordering: in the longer circuits included in our analysis, the former are approximately thrice that of the latter.  In these circumstances, the Jordan-Wigner and Bravyi-Kitaev mappings appear roughly equivalent for the smaller circuits. The Jordan-Wigner mapping outperforms the Bravyi-Kitaev mapping in some of the longer circuits, considering both total and entangling gate counts.  In the longest circuits considered (propanol), the use of the Jordan-Wigner mapping resulted in circuits that were approximately 25\% shorter.  This is attributed to the relative complexity of the Bravyi-Kitaev mapping resulting in a reduction of linearity in the CNOT chains, which hampers gate cancellation.

The error implications of Trotter ordering schemes are difficult to predict.  Bounds exist on the error yielded from Trotterization~\cite{poulin_trotter_2015}, although these are often loose~\cite{babbush_chemical_2015}.  A qualitative estimate can be obtained through determination of the norm of the Trotter error operator; however, the quantitative behaviour of this still often overestimates the error in real chemical examples.~\cite{babbush_chemical_2015}  It is useful to compare the implications of the Bravyi-Kitaev mapping when using several ordering schemes.  To this end, we repeated our calculations using four ordering schemes.  In addition to a single randomized ordering, a lexicographic ordering and an ordering of terms by magnitude, we include an ordering generated by regularly interspersing terms from the lexicographic and magnitude orderings.  Note that this ordering is relatively arbitrary and is intended for comparison purposes.  An exhaustive search of the ordering space is clearly intractable for non-trivial systems, owing to the factorial growth of the number of possible orderings.

Nonetheless, our findings are remarkably consistent, with the Bravyi-Kitaev mapping outperforming the Jordan-Wigner mapping in all cases apart from the lexicographic ordering.  This advantage increases with the number of spin-orbitals used.  Beyond $N=10$ in all non-lexicographic cases, the advantage associated with the Bravyi-Kitaev mapping exceeds that of just using gate cancellation.  This advantage is increased when considering CNOT count, and can be dramatic -- when using a magnitude ordering, the savings associated with the Bravyi-Kitaev mapping are approximately an order of magnitude greater than those obtained by using gate cancellation alone.

\begin{figure}[H]
	\centering
	\includegraphics[width=\aetfigsize]{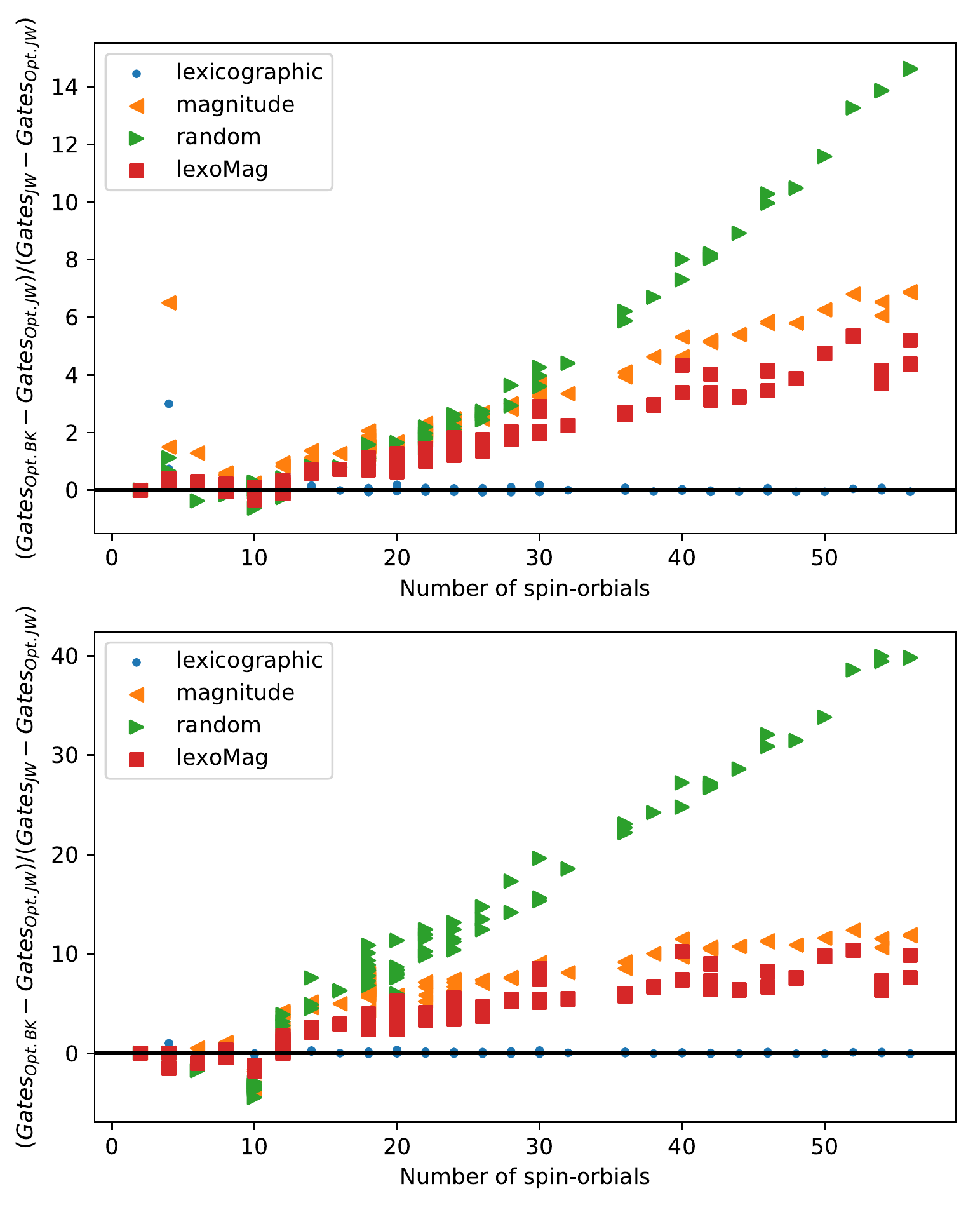}
	\caption{Gate savings associated with the Bravyi-Kitaev mapping normalised by the gate savings acquired using the same optimisation with the Jordan-Wigner mapping versus the number of spin-orbitals simulated, for various ordering schemes.  A value of 0 indicates that the optimized Bravyi-Kitaev and optimized Jordan-Wigner circuits have equal number of gates.  A value of 1 indicates that the optimized Bravyi-Kitaev mapping outperforms the optimized Jordan-Wigner mapping by a number of gates equal to that saved by performing optimization on the raw Jordan-Wigner circuit.  Upper: total gates.  Lower: entangling gates.}
	\label{plot:manyOrds}
\end{figure}

For large systems, searching even a statistically significant subset of the space of possible orderings is clearly intractable, owing to the factorial growth of the number of possibilities.  For each system, our random ordering is only one of these myriad choices.  Consequentially, it does not represent a statistically meaningful representation of the bulk of the possible orderings.  Nonetheless, it is interesting that this random choice qualitatively manifests the same trend as our other ordering schemes to a very large extent.  Quantitatively, the advantage associated with the Bravyi-Kitaev mapping when using these random orderings is dramatic - for our largest example, the reduction in CNOT count using the Bravyi-Kitaev mapping was forty times the reduction using gate cancellation alone.  In our results, the use of a tailored ordering scheme (whether it be ordering lexicographically, by magnitude or otherwise) results in a reduction of advantage for the Bravyi-Kitaev mapping.  

Although these figures suggest that the use of the Bravyi-Kitaev mapping results in shorter circuits when using most possible orderings, it should be emphasized that the lexicographic ordering dramatically decreases gate count through cancellation irrespective of mapping strategy.  We restate that using this strategy, Figure~\ref{plot:topLexGateComparison} shows that employing the Jordan-Wigner mapping results in marginally shorter circuits than the Bravyi-Kitaev mapping.  

The choice of Trotter ordering may be dictated by other factors - for instance, architecture constraints.  In these circumstances, calculations will be dramatically shortened by the use of the Bravyi-Kitaev mapping, as this mapping results in reduced circuit length for all of the non-lexicographic orderings considered.

The choice of ordering has previously been shown to hold significant impact on the Trotter error.~\cite{seeley_bravyi-kitaev_2012-5,poulin_trotter_2015}  Potentially, minimization of Trotter error may therefore require an ordering being chosen which is suboptimal in terms of single Trotter step gate count, where the Bravyi-Kitaev holds a significant advantage.

As such, consideration of the Bravyi-Kitaev mapping within the context of different ordering schemes requires an estimation of the associated Trotter error.  Previous work~\cite{seeley_bravyi-kitaev_2012-5,tranter_bravyi-kitaev_2015} indicates that for the hydrogen and methane molecules, the use of the Bravyi-Kitaev mapping can result in a reduced Trotter error - although, in the latter case, insufficiently to decrease the number of Trotter steps required for accurate simulation.  We consider these points in the general case in the following section.

In general, we conclude here that the use of the Bravyi-Kitaev mapping does not result in a predictable improvement in gate count when using an ordering optimized for gate cancellation, ignoring Trotter error implications.  This is in contrast with the substantial and predictable improvement observed with other orderings, particularly when a random ordering is used.

\section{Trotter Error Considerations}
\label{section:trotterError}
As discussed above, the use of Trotter-Suzuki approximations cause error which decreases with the number of Trotter steps performed.  Bounds on and estimates of this error have been established~\cite{poulin_trotter_2015}; however, it has been shown~\cite{babbush_chemical_2015} that these estimates often overestimate the actual error incurred by many orders of magnitude.  Exact determination of the Trotter error is exponentially hard, as the exact ground state energy must be determined in advance to serve as a reference.

Having generated Jordan-Wigner and Bravyi-Kitaev Hamiltonians as symbolic lists of Pauli strings, our code can proceed in several ways. For smaller systems, a sparse matrix representation of  these Hamiltonians can be generated using SciPy's~\cite{jones_scipy_2001} sparse matrix methods.  This can be diagonalised to find an exact ground-state eigenvalue (to compare against the estimate provided by further code) and a ground-state eigenvector.  These can be compared against traditional full configuration interaction calculations obtained through direct diagonalization of the Hamiltonian for verification purposes.

Our code can also generate Trotterized Jordan-Wigner and Bravyi-Kitaev Hamiltonians while maintaining the symbolic representation.  The action of these Hamiltonians on a given state can be simulated.  Performing this on the generated ground state eigenvectors allows for the determination of Trotter error without the storage difficulty of repeatedly generating the Trotterized evolution operator in matrix form.  Nonetheless, the initial determination of ground state eigenvectors remains exponentially difficult, requiring on the order of hundreds of gigabytes of RAM for examples with more than $20$ spin-orbitals.  Consequently, we restricted our error analysis to these smaller systems, as indicated by Table~\ref{table:systems1} and the appendices.

We conducted this analysis for 34 of our previously discussed systems.  The calculations were performed for a variety of choices of Trotter order and step number. Nonetheless, it is remarkable that limiting our discussion to trials of one Trotter step of first order suffices to yield chemical accuracy (i.e. to within 1 kcal/mol of the FCI energy) for a single application of the evolution operator.  For simulation of the full phase estimation procedure, determination of the error incurred in the application of higher powers of this operator would be necessary.  However, this does not qualitatively effect our ordering comparison.

\begin{figure}[h!]
	\centering
	\includegraphics[width=\aetfigsize]{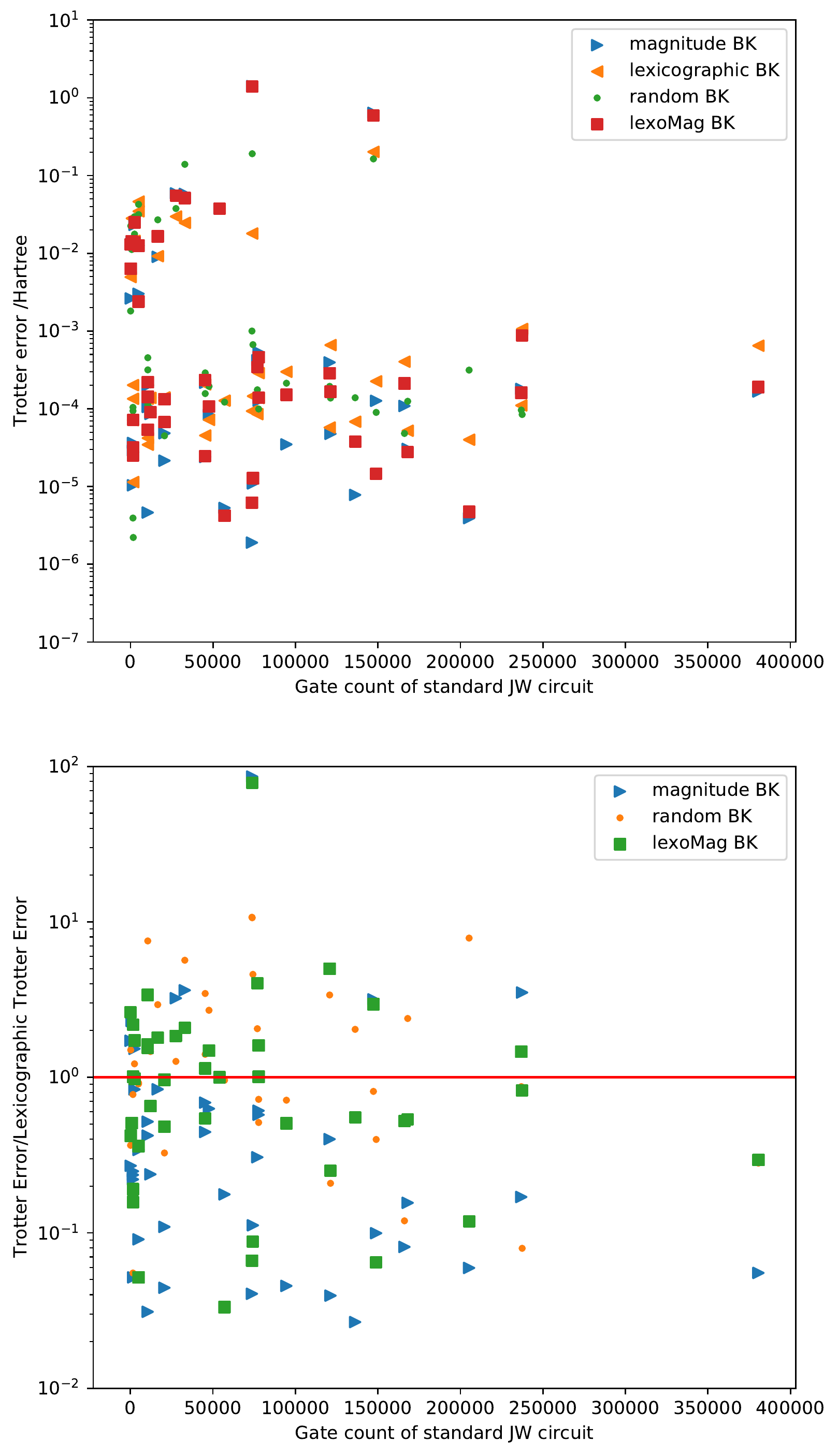}
	\caption{Trotter error using the Bravyi-Kitaev mapping as a function of Jordan-Wigner circuit length, for different ordering schemes.  Upper: Absolute Trotter error.  Middle: Trotter error relative to error of lexicographic ordering.  Note this excludes two instances where the relative Trotter error was $>$ 20. The red line indicates a relative performance of 1 - i.e. below the line, the ordering results in a lower Trotter error than a lexicographic ordering.  Note that the magnitude ordering usually results in a substantially lower error, however in most of these cases the Trotter error was already around $10^{-4}$.}
	\label{plot:trotterErrors1}
\end{figure}

Figure~\ref{plot:trotterErrors1} demonstrates the results of these simulations.  Encouragingly, in the overwhelming majority of systems considered, the Trotter error is extremely small even with only one Trotter step: it is often less than 0.001 Hartree. It is possible that this is an artefact of the small number of spin orbitals in the systems considered in our error analysis.  It is also worth noting that these errors will be compounded when considering higher bits of precision in a full phase estimation procedure.  Nonetheless, it does suggest that the number of Trotter steps required for chemically accurate simulation of larger systems will be relatively modest, potentially less than $10$ Trotter steps for the first bit of precision.

Figure~\ref{plot:trotterErrors1} additionally demonstrates the ordering dependence of the Trotter error.  We consider the Trotter error of each systematic (i.e. non-random) ordering normalised by the Trotter error of the lexicographic ordering for each system.

In most cases, the magnitude ordering appears to achieve a lower Trotter error than the lexicographic ordering.  In some cases, this difference exceeds an order of magnitude.  However, it is important to note that this represents a large variance on an exceptionally small error.  Noting that one Trotter step was sufficient for chemical accuracy in most of the systems studied here, we do not argue that this indicates that a magnitude ordering achieves a useful reduction in error compared to a lexicographic ordering.  Future work is required to investigate how significant this distinction is when propagated through the entire phase estimation procedure, as in these circumstances this effect could become sufficiently significant to determine ordering choice.  

\begin{figure}[h!]
 	\centering
 	\includegraphics[width=\aetfigsize]{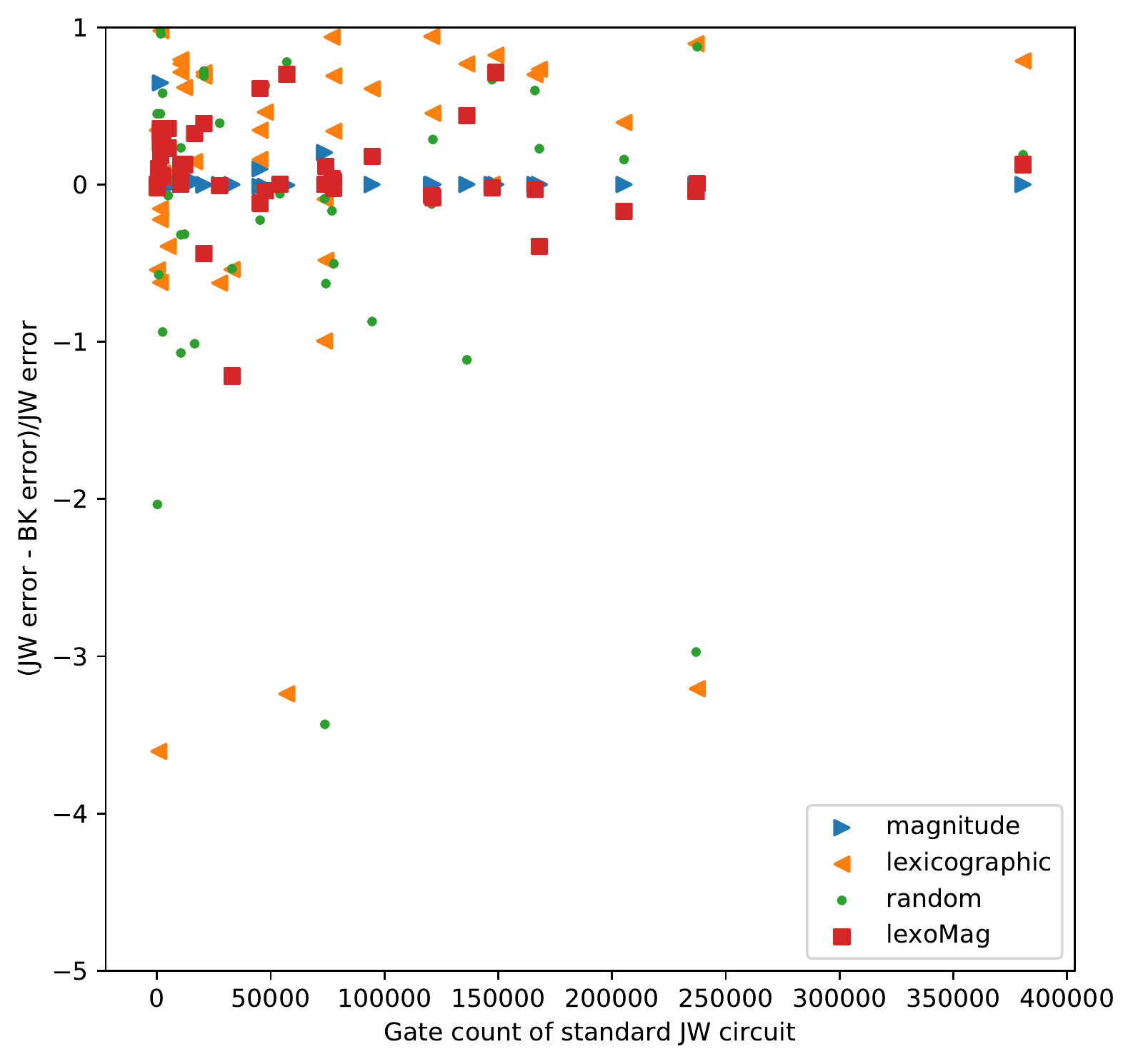}
 	\caption{Relative Trotter error of Jordan-Wigner and Bravyi-Kitaev mappings as a function of gate count.  The difference in error is normalised by the Jordan-Wigner error, such that a value of 0 indicates equivalent performance, with negative values implying high Bravyi-Kitaev error.  Both schemes display remarkably similar errors for the magnitude ordering, likely due to the high degree of ``physicality" of the ordering.  For a lexicographic ordering, the Bravyi-Kitaev mapping shows lower error in most of the systems studied.}
 	\label{plot:trotterErrors2}
 \end{figure}

 An examination of the relative performance of the Bravyi-Kitaev mapping and the Jordan-Wigner mapping in the context of ordering strategies is included as Figure~\ref{plot:trotterErrors2}.  Again, the distinction between the two mappings is in most cases not as substantial as the differences observed for single Trotter step circuit length.  In the majority of systems, the normalised difference between the two errors is between 1 and -1 -- that is, the error associated with one mapping is very rarely more than double that of the other.  Several systems display substantially increased error associated with the Bravyi-Kitaev mapping (including two not shown on Figure ~\ref{plot:trotterErrors2} for scale clarity); however, in the general case no such pattern emerges.
 Using a lexicographic ordering, a preference for the Bravyi-Kitaev mapping is observable in most systems.  In some cases, the error is almost halved by the use of the Bravyi-Kitaev mapping.  At larger system sizes this could become a more substantial effect; however, we do not contend that our data provides concrete evidence as to whether this is true.  For a magnitude ordering the Jordan-Wigner and Bravyi-Kitaev mappings yield nearly identical Trotter error in almost all cases.   In a sense, this could be attributed to the more directly physical nature of the magnitude ordering.  Important terms will intrinsically be simulated earlier in sequence using both mappings.  As such, the error of both is likely to be similar in this case.  As to be expected, the ``lexoMag" ordering performs roughly as a combination of the magnitude and lexicographic orderings.

As above, the impact of this variation in error could prove substantial when propagated through the entire phase estimation algorithm.  Using a lexicographic ordering -- optimal for gate cancellation -- the Bravyi-Kitaev mapping outperforms the Jordan-Wigner mapping in most cases.  If this superior error performance scales to above 30 qubit systems, this could result in a reduction of the necessary Trotter steps for simulation.  The consequent reduction in circuit length could counterbalance the marginally increased individual Trotter step cost of using the Bravyi-Kitaev mapping in a lexicographic ordering.  Further work examining the entire procedure should assess this.  Nonetheless, it should not be forgotten that an exact determination of the Trotterization error is equivalent to the solution of the exponentially hard eigenvalue problem itself.  Consequentially, this approach may prove intractable.  Qualitatively, an examination of the norm of the Trotter error operator may prove instructive.~\cite{babbush_chemical_2015}

\section{Further Optimized Circuits}
\label{section:further}
\begin{figure}
	\centering
	\includegraphics[width=\aetfigsize]{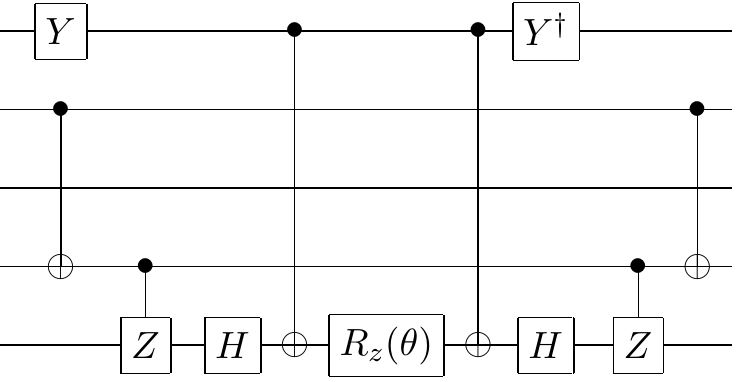}
	\caption{A circuit performing an equivalent operation to Figure ~\ref{plot:circ1}, using a basis change shift optimisation.}
	\label{plot:circ3}
\end{figure}

	Our code also generates some of the optimised circuits developed by Hastings, Wecker, Bauer \& Troyer~\cite{hastings_improving_2015}, in order to assess the impact of these optimisations with respect to the Bravyi-Kitaev mapping.  Examining Figure ~\ref{plot:circ1}, it is evident that many of the CNOT strings may be ``blocked off" from cancellation by the basis change gates exterior to them.  In this approach, these basis change gates are brought inside the bulk of the CNOT string, as shown in Figure ~\ref{plot:circ3}.  Note that in the first CNOT string, the final CNOT is replaced by a CZ gate.  An inspection of the gate sequence implemented on the final qubit in the case of even and odd parities of each subset of qubits demonstrates why this is the case, as discussed in the appendices.

Our implementation is a slight modification of this technique.  As discussed above, our code reduces the Hamiltonian to a symbolic list of exponentiated Pauli strings which does not preserve the electronic Hamiltonian's original $H_{pqrs}$ components.  In this scheme, it is not always the case that the ``final" qubit - the qubit which is acted on by the central single qubit rotation - is always in the $X$ basis.  As such, the above method requires modification.  For example, if the central qubit is to be rotated in the $Z$ basis, the additional CZ gate could simply be commuted through the central rotation and cancelled, leading to a phase error.

Fortuitously, in the case that the central qubit is to be rotated in either the $Y$ or $Z$ basis, a CNOT gate can be used in place of the CZ gate, as in Figure ~\ref{plot:circ3}.  To demonstrate this, we consider the action on the final qubit in the case of the parity of each subset of qubits, as shown in the appendices.

Additionally, circuits described by Hastings, Wecker, Bauer \& Troyer~\cite{hastings_improving_2015} involving an ancilla qubit can be generated.  Here, all parity calculating CNOT gates are targeted on a single ancilla qubit.  This allows for CNOTS performed in the same basis to be moved around arbitrarily, allowing for a greater level of gate cancellation.  

Circuits of these forms are known to reduce overall gate count substantially.~\cite{hastings_improving_2015}  We focus here on the relevance of the Bravyi-Kitaev mapping when using these techniques.

Using the former technique, the performance of the Bravyi-Kitaev technique using a lexicographic or random ordering displays roughly the same trend as previous circuits.  However, using a magnitude ordering reduces the efficacy of the Bravyi-Kitaev mapping to the point of near-equivalence to the Jordan-Wigner mapping.  Any savings or penalties associated with the use of the Bravyi-Kitaev mapping are then negligible compared to gains from the basis change shift optimisation procedure. We do not yet have an explanation for the ordering dependence of this behaviour.  Nonetheless, the Bravyi-Kitaev mapping never substantially under-performs when compared to the Jordan-Wigner mapping.  In essence, the observed trends are the same as for the previous circuits, albeit with a greatly reduced factor of improvement associated with the use of the Bravyi-Kitaev mapping.
\begin{figure}[h!]
	\centering
	\includegraphics[width=\aetfigsize]{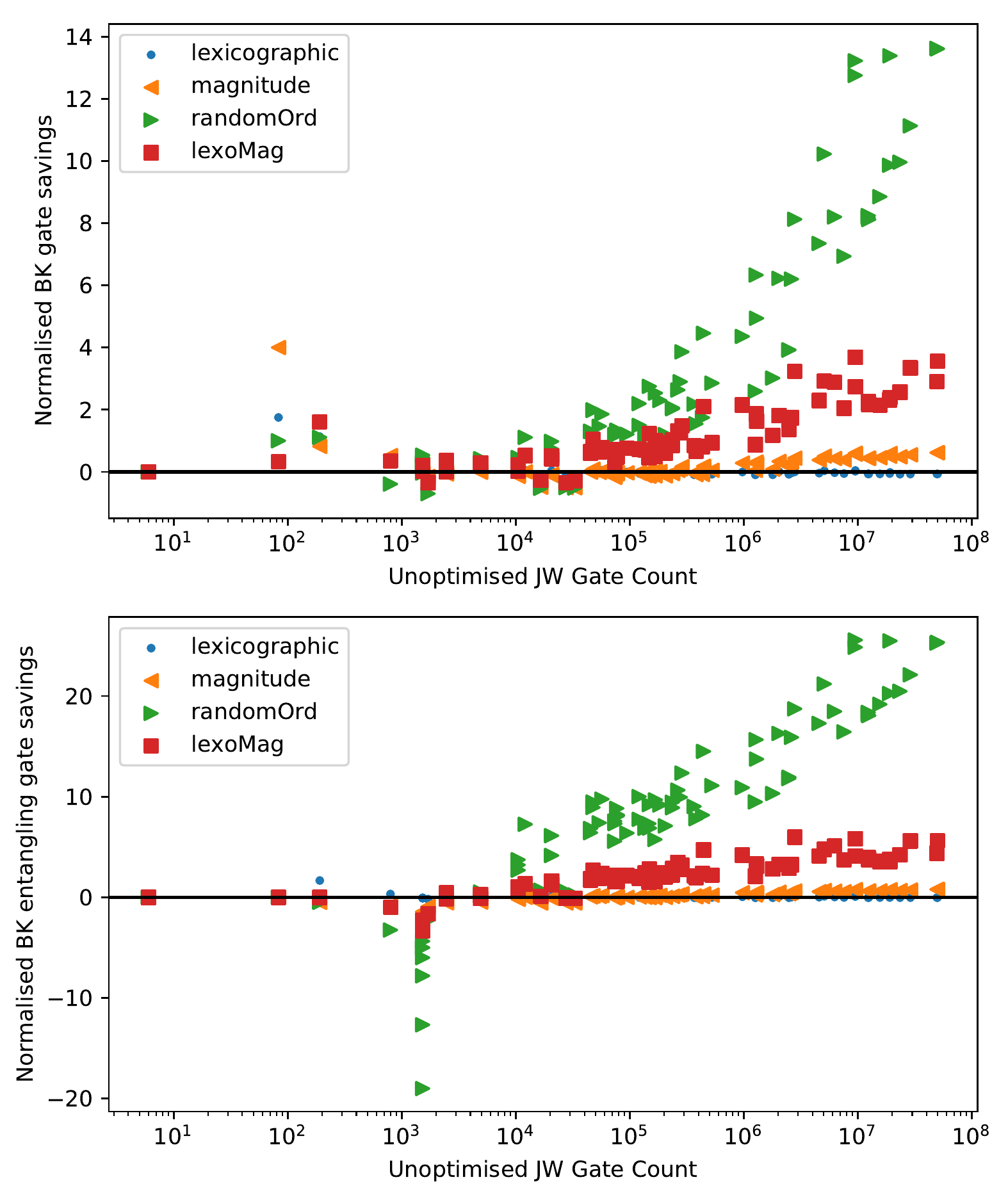}
	\caption{Gate savings associated with the Bravyi-Kitaev mapping normalised by the gate savings acquired using the same optimisation with the Jordan-Wigner mapping, using the modified basis change shift technique. Upper:  Total gate count.  Lower:  Entangling gate count.  Here we see that the Bravyi-Kitaev and Jordan-Wigner mappings perform relatively equivalently when using both lexicographic and magnitude orderings.  Using an alternative ordering, the Bravyi-Kitaev mapping is superior, however these result in an overall increased gate count in both cases.}
	\label{plot:futherOps}
\end{figure}

The use of ancilla circuits displays a markedly different trend.  Here, the effect of the Bravyi-Kitaev mapping is greatly reduced in all ordering schemes -- there is a maximum of around 30\% reduction in the largest examples, when using a ``lexoMag" ordering.  It is likely that any major savings or penalties associated with the Bravyi-Kitaev mapping are masked by the substantial savings associated with the use of ancillarised circuits.  Using a lexicographic ordering, there is no predictable difference between the two mapping schemes at all.  

Curiously, using a magnitude ordering reverses the trend observed for previous optimization schemes.  Here, the Bravyi-Kitaev mapping is consistently outperformed by the Jordan-Wigner mapping.  However, this distinction is relatively small, and disappears entirely in larger system sizes.  As such, we do not conclude that a consistent preference for either mapping is present using these circuits.  Note that the use of such circuits may be undesirable in certain architectures, due to the loss of nearest-neighbour CNOT chains, which could undermine the substantial savings associated with this technique.
\begin{figure}[h!]
	\centering
	\includegraphics[width=\aetbigfigsize]{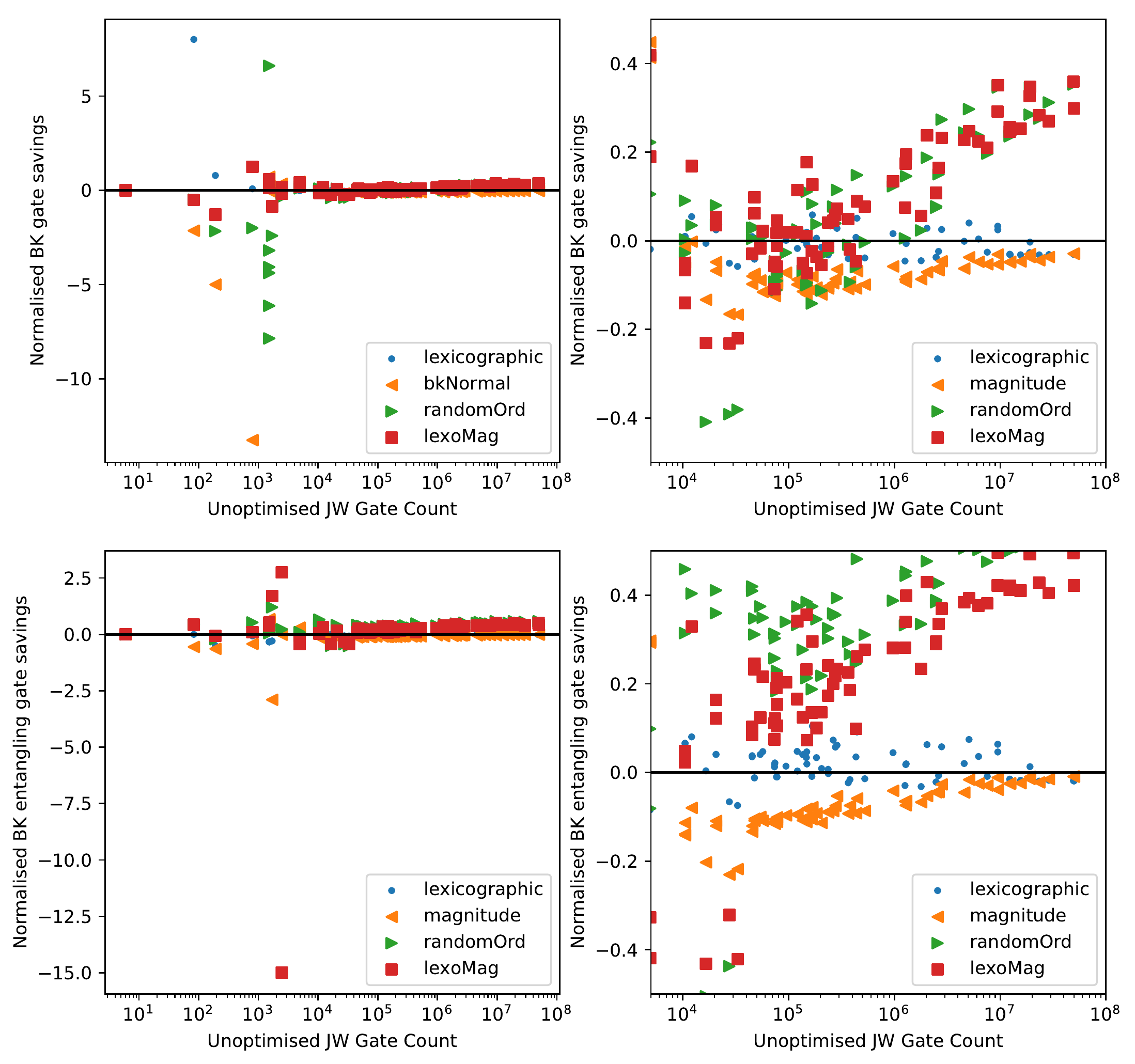}
	\caption{Gate savings associated with the Bravyi-Kitaev mapping normalised by the gate savings acquired using the same optimisation with the Jordan-Wigner mapping, using the ancilla-based technique. Upper left:  Total gate count.  Upper right: Total gate count, zoomed.  Lower left:  Entangling gate count. Lower right: Entangling gate count, zoomed.  Here, we again see roughly equivalent performance when using either a magnitude or lexicographic ordering.  Once again, when using a alternative ordering, the Bravyi-Kitaev mapping is superior, however to a greatly reduced extent.}
	\label{plot:futherOps}
\end{figure}

Finally, we note that in all optimisation systems studied, a random choice of ordering resulted in the strongest performance of the Bravyi-Kitaev mapping.  While one ordering is clearly not a statistically meaningful sample of the possible orderings, it is interesting that our results here are extremely consistent.

\section{Conclusions}
In this paper we have made a detailed comparison of the Jordan-Wigner and Brayvi-Kitaev mappings using a variety of advanced circuit optimisation techniques drawn from the theory of quantum simulation.  Using unoptimized circuits, the use of the Bravyi-Kitaev mapping dramatically reduces the quantum computational expense of simulation in all systems involving more than 30 qubits, and for systems likely to be classically intractable to simulate.  In cases with approximately 50 qubits, this improvement reduced the gate count by roughly 25\%.

The use of an optimised Trotter ordering absorbs the advantage associated with the Bravyi-Kitaev mapping.  The Jordan-Wigner mapping in this case results in slightly shorter circuits for an individual Trotter step.  Nonetheless, in most cases the use of the Bravyi-Kitaev mapping is at worst roughly equivalent to the use of the traditional Jordan-Wigner mapping.  Frequently, the gate count reduction is particularly large in the number of expensive entangling gates required.  Notably, the Bravyi-Kitaev mapping is superior in all cases aside from the lexicographic ordering.

While our results suggest a slightly reduced error associated with the use of a magnitude ordering, we do not conclude that this ordering should be favoured due to the substantial overall gate count associated with a lexicographic ordering.  Our analysis of Trotter ordering error suggests that the use of the Bravyi-Kitaev mapping typically results in a reduced Trotter error when using any ordering studied, other than a magnitude ordering.  This difference is almost a factor of 2 in many larger examples when using a lexicographic ordering.  This encourages the use of the Bravyi-Kitaev mapping, as it could outweigh the relatively small benefit from the use of the Jordan-Wigner mapping in optimised lexicographic circuits.

Our results demonstrate that the performance of the Bravyi-Kitaev mapping is dependent on a variety of factors.  While it is superior to the Jordan-Wigner mapping in most cases studied, several exceptions were observed.  This emphasizes the importance of numerical analysis in future work.  It is apparent that such studies must be performed across a range of molecular systems, with due consideration given to the region where classical full configuration interaction calculations are intractable.
\color{black}

  Recent hardware developments have prompted the suggestion that quantum devices could be used for practical tasks in as little as five years.~\cite{mohseni_commercialize_2017}  The use of quantum computers to perform classically intractable quantum chemistry calculations is often cited as one of the principal uses of emerging quantum technology.~\cite{olson_quantum_2017}

We have demonstrated here that the use of the Bravyi-Kitaev transformation frequently results in substantially reduced gate count estimates.  In the future, we anticipate that the application of this mapping will assist in the performance of electronic structure calculations on real quantum computers, yielding results that have proven computationally elusive for classical devices.

	\clearpage
\section*{Author Information}
(AT) E-mail: a.tranter13@imperial.ac.uk

(PVC) E-mail: p.v.coveney@ucl.ac.uk

\subsubsection*{ORCID}

Peter V. Coveney: 0000-0002-8787-7256

\subsubsection*{Funding}

AT acknowledges EPSRC for a PhD Studentship through Imperial College London's Centre for Doctoral Training in Controlled Quantum Dynamics (grant number EP/G037043/1).  PVC thanks EPSRC for funding via grant number EP/L00030X/1.  PJL acknowledges support by AFOSR award no. FA9550-12-1-0046.

\subsubsection*{Notes}

The authors declare no competing financial interest.

\begin{acknowledgement}
 The authors thank Jarrod McClean and Ryan Babbush for providing molecular orbital integral data used for preliminary versions of this analysis.

\end{acknowledgement}

\bibliography{BK_paper}

\begin{mcitethebibliography}{48}
\bibitem[Not()]{Note-1}
This requirement can be relaxed by additional encoding
  techniques,~\cite{bravyi_tapering_2017,steudtner_lowering_2017} however we do
  not consider these techniques here.\relax
\mciteBstWouldAddEndPunctfalse
\mciteSetBstMidEndSepPunct{\mcitedefaultmidpunct}
{}{\mcitedefaultseppunct}\relax
\EndOfBibitem
\bibitem[Not()]{Note-2}
This iterative process is not optimised for classical computing cost, and would
  likely be performed more efficiently when generating circuits to be
  implemented on real quantum devices.\relax
\mciteBstWouldAddEndPunctfalse
\mciteSetBstMidEndSepPunct{\mcitedefaultmidpunct}
{}{\mcitedefaultseppunct}\relax
\EndOfBibitem
\end{mcitethebibliography}
\begin{appendices}
	\section{List of systems studied}
	\csvstyle{systemsTableStyle}{longtable=lccccc,
		table head=\caption{All molecules considered in this study. Level corresponds to whether or not the error incurred in Trotterization was considered -- N indicates that this analysis was not performed for this system, Y indicates that it was.}\\\toprule\bfseries Name & \bfseries Charge & \bfseries Mult. & \bfseries Basis & \bfseries N. Qubits & \bfseries Level \\\midrule,
		late after line=\\,
		late after last line=\\\bottomrule,
		head to column names}
	
	\begin{filecontents*}{molecules.csv}
molname,charge,multiplicity,basis,qubits,error
Methane,0,1,STO-3G,18,Y
Ethane,0,1,STO-3G,32,N
Propane,0,1,STO-3G,46,N
Ethene,0,1,STO-3G,28,N
Propene,0,1,STO-3G,42,N
Ethyne,0,1,STO-3G,24,N
Methanol,0,1,STO-3G,28,N
Ethanol,0,1,STO-3G,42,N
Isopropanol,0,1,STO-3G,56,N
1-Propanol,0,1,STO-3G,56,N
Propanone,0,1,STO-3G,52,N
Methanal,0,1,STO-3G,24,N
Ethanal,0,1,STO-3G,38,N
Ethanoate Ion,-1,1,STO-3G,46,N
Ethanoic Acid,0,1,STO-3G,48,N
Hydrogen Peroxide,0,1,STO-3G,24,N
Ethanamide,0,1,STO-3G,50,N
Methylamine,0,1,STO-3G,30,N
Dimethylamine,0,1,STO-3G,44,N
Ammonia,0,1,STO-3G,16,Y
Ammonium,1,1,STO-3G,18,Y
Nitrogen Dioxide,0,2,STO-3G,30,N
Lithium Hydroxide,0,1,STO-3G,22,Y
Sodium Hydroxide,0,1,STO-3G,30,N
H\textsubscript{2},0,1,STO-3G,4,Y
Lithium Hydride,0,1,STO-3G,12,Y
Beryllium Hydride,0,1,STO-3G,14,Y
N\textsubscript{2},0,1,STO-3G,20,Y
O\textsubscript{2},0,3,STO-3G,20,Y
O\textsubscript{2},0,1,STO-3G,20,Y
F\textsubscript{2},0,1,STO-3G,20,Y
Sodium Hydride,0,1,STO-3G,20,Y
Magnesium Hydride,0,1,STO-3G,22,Y
Cl\textsubscript{2},0,1,STO-3G,36,N
HCl,0,1,STO-3G,20,Y
HF,0,1,STO-3G,12,Y
Carbon Dioxide,0,1,STO-3G,30,N
Carbon Monoxide,0,1,STO-3G,20,Y
Water,0,1,STO-3G,14,Y
Methylene,0,3,STO-3G,14,Y
Hydroxide,-1,1,STO-3G,12,Y
HNO\textsubscript{3},0,1,STO-3G,42,N
Nitrate,-1,1,STO-3G,40,N
H,0,2,STO-3G,2,N
He,0,1,STO-3G,2,N
Li,0,2,STO-3G,10,Y
Be,0,1,STO-3G,10,Y
B,0,2,STO-3G,10,Y
C,0,3,STO-3G,10,Y
N,0,4,STO-3G,10,N
O,0,3,STO-3G,10,N
F,0,2,STO-3G,10,N
Ne,0,1,STO-3G,10,N
Na,0,2,STO-3G,18,Y
Mg,0,1,STO-3G,18,Y
Si,0,3,STO-3G,18,Y
P,0,4,STO-3G,18,Y
S,0,3,STO-3G,18,Y
Cl,0,2,STO-3G,18,Y
Ar,0,1,STO-3G,18,N
K,0,2,STO-3G,26,N
I,0,2,STO-3G,54,N
Br,0,2,STO-3G,36,N
I\textsuperscript{-},-1,1,STO-3G,54,N
Br\textsuperscript{-},-1,1,STO-3G,36,N
Cl\textsuperscript{-},-1,1,STO-3G,18,N
H\textsubscript{2},0,1,3-21G,8,Y
H\textsubscript{2},0,1,6-31G,8,Y
H\textsubscript{2},0,1,6-31G**,20,Y
H\textsubscript{2},0,1,6-311G*,12,Y
H\textsubscript{2},0,1,6-311G**,24,N
HeH\textsuperscript{+},1,1,STO-3G,4,Y
HeH\textsuperscript{+},1,1,3-21G,8,Y
HeH\textsuperscript{+},1,1,6-31G,8,Y
HeH\textsuperscript{+},1,1,6-31G**,20,Y
HeH\textsuperscript{+},1,1,6-311G*,12,Y
HeH\textsuperscript{+},1,1,6-311G**,24,N
Methylene,0,3,3-21G,14,N
HF,0,1,3-21G,22,Y
Lithium Hydride,0,1,3-21G,22,Y
H\textsubscript{2}O,0,1,3-21G,26,N
H3\textsuperscript{+},1,1,STO-3G,6,Y
H3\textsuperscript{+},1,1,3-21G,12,Y
CO\textsubscript{3},0,1,STO-3G,40,N
Magnesium Hydroxide,0,1,STO-3G,42,N
H\textsubscript{2}S,0,1,STO-3G,22,Y
\end{filecontents*}
\csvreader[systemsTableStyle]
	{molecules.csv}{}{\molname & \charge & \multiplicity & \basis & \qubits & \error}%
	\clearpage
	\section{Basis change shift}
	The following tables indicate the operation which is performed upon a qubit having modified the original circuit in line with section ~\ref{section:further}.  Note that the end results of each modified circuit is equivalent to that of the original circuit.
	
		\begin{tabular}[h]{|l|r|}
		\hline
		\multicolumn{2}{|c|}{Original Circuit, Central Z}\\
		\hline
		Parity& Central Gate Sequence  \\
		\hline
		Even & $R_z$  \\
		Odd & $X R_z X$  \\
		\hline
	\end{tabular}
\\
\\
		\begin{tabular}[h]{|l|c|c|r|}
			\hline
			\multicolumn{4}{|c|}{Modified Circuit, Central Z}\\
			\hline
			Exterior Parity& Interior Parity &  Overall Parity & Central Gate Sequence  \\
			\hline
			Even & Even & Even & $R_z$  \\
			Even & Odd & Odd & $X R_z X$  \\
			Odd & Even & Odd & $X R_z X$  \\
			Odd & Odd & Even & $XX R_z XX = R_z$  \\
			\hline
		\end{tabular}	
		
		\begin{tabular}{|l|r|}
			\hline
			\multicolumn{2}{|c|}{Original Circuit, Central Y}\\
			\hline
			Parity& Central Gate Sequence  \\
			\hline
			Even & $Y R_z Y^\dagger$  \\
			Odd & $Y X R_z X Y^\dagger$  \\
			\hline
		\end{tabular}
		
		\begin{tabular}{|l|c|c|r|}
			\hline
			\multicolumn{4}{|c|}{Modified Circuit, Central Z}\\
			\hline
			Exterior Parity& Interior Parity &  Overall Parity & Central Gate Sequence  \\
			\hline
			Even & Even & Even & $Y R_z Y^\dagger$  \\
			Even & Odd & Odd & $Y X R_z X Y^\dagger$  \\
			Odd & Even & Odd & $X Y R_z  Y^\dagger X= Y X R_z X Y^\dagger$  \\
			Odd & Odd & Even & $X Y X R_z X Y^\dagger X =Y R_z Y^\dagger$   \\
			\hline
		\end{tabular}

	\end{appendices}
	\end{document}